\documentclass[10pt]{article}
\usepackage{graphicx, amsmath, amssymb, amsfonts, amsthm, authblk, listings, color, hyperref, longtable}

\definecolor{mygreen}{rgb}{0,0.6,0}
\definecolor{mygray}{rgb}{0.85,0.85,0.85}
\definecolor{mymauve}{rgb}{0.58,0,0.82}

\newcommand{\py}{\lstinline[language=Python]}

\title{Who is the Centre of the Movie Universe?\\\smallskip
\large Using Python and NetworkX to Analyse the Social Network of Movie Stars}
\author{Rhyd Lewis}
\affil{
	School of Mathematics,\\
	Cardiff University, Cardiff, Wales.\\
	\url{LewisR9@cf.ac.uk}, \url{http://www.rhydlewis.eu}
}

\begin{document}
\maketitle

\abstract{This paper provides the technical details of an article originally published in \emph{The Conversation} in February 2020~\cite{Lewis2020Conv}. The purpose is to use centrality measures to analyse the social network of movie stars and thereby identify the most ``important'' actors in the movie business. The analysis is presented in a step-by-step, tutorial-like fashion and makes use of the Python programming language together with the NetworkX library. It reveals that the most central actors in the network are those with lengthy acting careers, such as Christopher Lee, Nassar, Sukumari, Michael Caine, Om Puri, Jackie Chan, and Robert De~Niro. We also present similar results for the movie releases of each decade. These indicate that the most central actors since the turn of the millennium include people like Angelina Jolie, Brahmanandam, Samuel L. Jackson, Nassar, and Ben Kingsley.}

%---------------------------------------------------------------------------
\section{Introduction}

Social network analysis is a branch of data science that allows the investigation of social structures using networks and graph theory. It can help to reveal patterns in voting preferences, aid the understanding of how ideas spread, and even help to model the spread of diseases~\cite{wiki,Needham2019,Wasserman1994}. 

A social network is made up of a set of \emph{nodes} (usually people) that have links, or \emph{edges} between them that describe their relationships. In this article we analyse the social network formed by movie actors. Each actor in this network is represented as a node. Pairs of actors are then joined by an edge if they are known to have appeared in a movie together. This information is taken from the Internet Movie Database IMDb~\cite{IMDb}. Our analysis is carried out using the Python programming language and, in particular, the tools available in the \emph{NetworkX} library~\cite{NetworkX}.

\section{A Small Example} 

\begin{figure}
\begin{center}
\fbox{\includegraphics[scale=0.33, clip=true, trim=0 50 270 0]{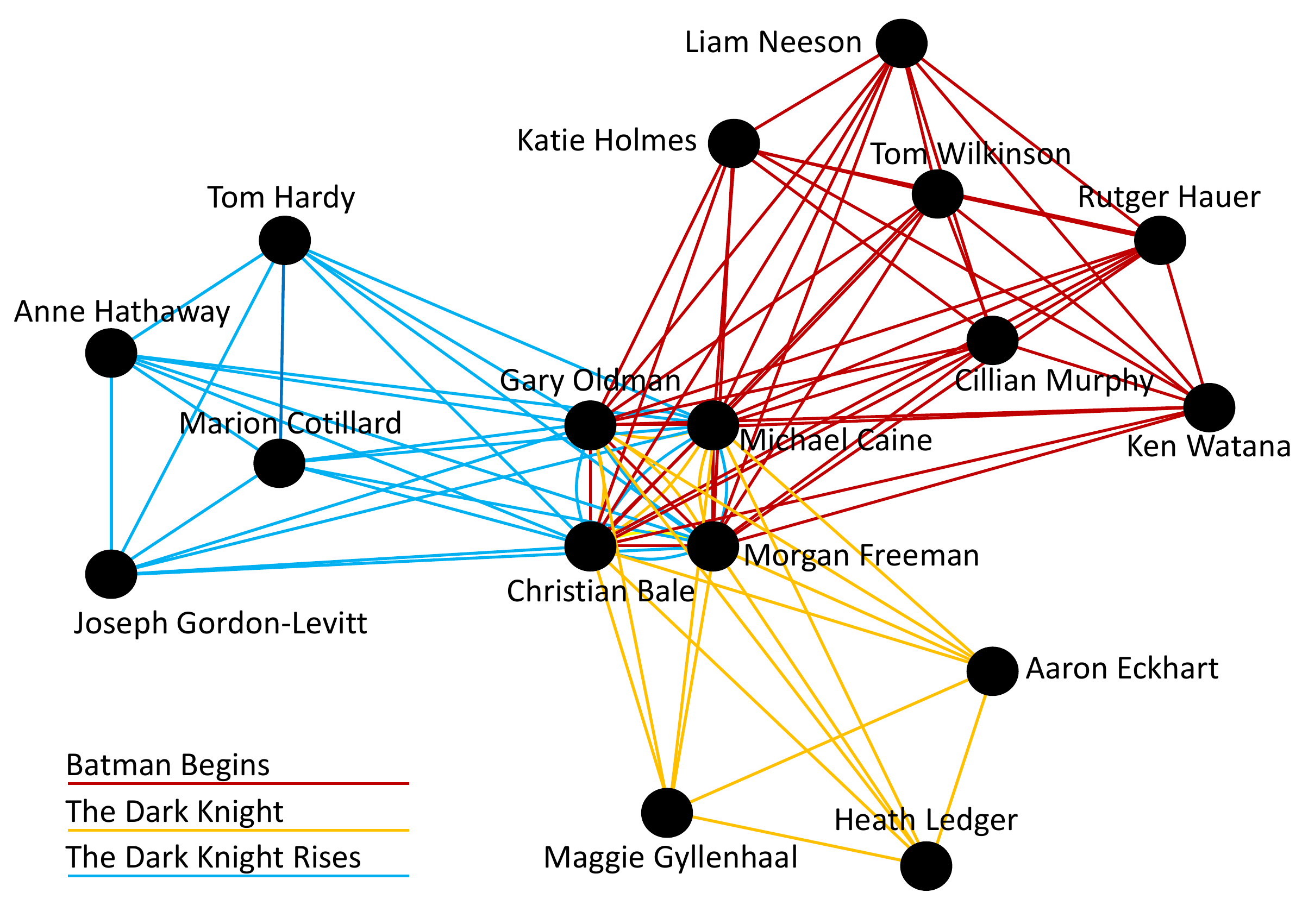}}
\caption{Relationships between actors appearing in \emph{The Dark Knight Trilogy}.}
\label{fig:batmanNetwork}
\end{center}
\end{figure}

Figure~\ref{fig:batmanNetwork} shows a small social network formed by the actors appearing in Christopher Nolan's three Batman movies, \emph{The Dark Knight Trilogy}. As mentioned, each node in this network corresponds to an individual actor. An edge between a pair of nodes then indicates that the two actors appeared in the same movie together. 

A number of features are apparent in this network. We can see that the nodes seem to be clustered into four groups. The tight cluster in the centre contains Christian Bale, Michael Caine, Gary Oldman and Morgan Freeman, who starred in all movies of the trilogy. In contrast, the remaining clusters hold the actors who appeared in just one of the movies. The cluster at the top-right shows the actors who appeared in \emph{Batman Begins}, the cluster at the bottom contains the stars of \emph{The Dark Night}, and the cluster on the left shows the actors from \emph{The Dark Night Rises}. We also see, for example, that Tom Hardy was in the same movie as Joseph Gorden Levitt (in \emph{The Dark Knight Rises}), but did not appear alongside actors such as Liam Neeson (who was a star of \emph{Batman Begins}), or Heath Ledger (who appeared in \emph{The Dark Night}).

%---------------------------------------------------------------------------
\section{A Dataset of All Movies}

While the Batman example shown in Figure~\ref{fig:batmanNetwork} is helpful for illustrative purposes, in this article we are interested in investigating the social network of \emph{all} actors from \emph{all} movies. As mentioned, for this study we use information taken from the Internet Movie Database~\cite{IMDb}. Specifically, we use a dataset compiled by the administrators of the Oracle of Bacon website~\cite{OoB}. Complete and up-to-date versions of this dataset can be downloaded directly from~\cite{data}. 

Our version of this dataset was downloaded at the start of January 2020 and contains the details of 164,318 different movies. Each movie in this set is stored as a JSON object containing, among other things, the title of the movie, a list of the cast members, and the year of its release. The complete dataset it is obviously too large to reproduce here, but to illustrate the basic format, the box below shows the three-movie example used to produce the small social network shown in Figure~\ref{fig:batmanNetwork}.

\begin{lstlisting}[frame=single]
{"title":"Batman Begins","cast":["Christian Bale","Michael Caine","Liam Neeson","Katie Holmes","Gary Oldman","Cillian Murphy","Tom Wilkinson","Rutger Hauer","Ken Watanabe","Morgan Freeman"],"year":2005}
{"title":"The Dark Knight","cast":["Christian Bale","Michael Caine","Heath Ledger","Gary Oldman","Aaron Eckhart","Maggie Gyllenhaal","Morgan Freeman"],"year":2008}
{"title":"The Dark Knight Rises","cast":["Christian Bale","Michael Caine","Gary Oldman","Anne Hathaway","Tom Hardy","Marion Cotillard","Joseph Gordon-Levitt","Morgan Freeman"],"year":2012}
\end{lstlisting}

Before proceeding with our analysis, note that is was first necessary to remove a few ``dud'' movies from this dataset. In our case, we decided to remove the 44,075 movies that had no cast specified. We also deleted a further 5,416 movies that did not include a year of release. This leaves a final ``clean'' database of 114,827 movies with which to work. In the following Python code we call this file \py{data.json}.

%---------------------------------------------------------------------------
\section{Input and Preliminary Analysis}

In this section we show how the dataset can be read into our program using standard Python commands. We then carry out a preliminary analysis of the data, produce some simple visualisations, and use these to help identify some inconsistencies in the dataset.

\subsection{Reading the Dataset}

To read the dataset, we begin by first importing the relevant Python libraries into our program. Next, we transfer the contents of the entire dataset into a Python list called \py{Movies}. Each element of this list contains the information about a single movie. The command \py{json.loads(line)} is used to convert each line of raw text (in JSON format) into an appropriate Python data structure. This is then appended to the \py{Movies} list. 

\begin{lstlisting}[language = python, frame=single]
import json
import networkx as nx
import matplotlib.pyplot as plt
import collections
import statistics 
import time
import random

Movies = []
with open("data.json", "r", encoding="utf-8") as f:
    for line in f.readlines():
        J = json.loads(line)
        Movies.append(J)
\end{lstlisting}

Having parsed the dataset in this way, we are now able to access any of its elements using standard Python commands. For example, the statement \py{Movies[0]} will return the full record of the first movie in the list; the statement \py{Movies[0]["cast"][0]} will return the name of the first cast member listed for the first movie; and so on.

\subsection{Two Simple Bar Charts}

Having read the dataset into the list \py{Movies}, we can now carry out some basic analysis. Here we will look at the number of movies produced per year, and the sizes of the casts that were used. The code below uses the \py{collections.Counter()} method to count the number of movies released per year. This information is written to the variable \py{C}, which is then used to produce a bar chart via the \py{plt.bar()} command. 

\begin{lstlisting}[language = python, frame=single]
C = collections.Counter([d["year"] for d in Movies])
plt.xlabel("Year")
plt.ylabel("Frequency")
plt.title("Number of Movies per Year")
plt.bar(list(C.keys()), list(C.values()))
plt.show()
\end{lstlisting}

The resultant bar chart is shown below. As we might expect, we see that nearly all movies in this dataset were released between the early 1900's and 2020, with a general upwards trend in the number of releases per year. However, the fact that the horizontal axis of our chart goes all the way back to 1800 hints at the existence of outliers and errors in the dataset. In fact, a few errors do exist. For example, the movie \emph{Cop} starring James Woods is stated as being released in the year 1812, which is clearly ridiculous (James Woods wasn't born until 1947, and \emph{Cop} was actually released in 1988). On the other hand, a movie called \emph{Avatar 5} is given a ``release date'' of 2025 in the dataset, which is also incorrect (at present, only one \emph{Avatar} movie has been made). Nevertheless, we will accept such oddities and continue with our investigation. 

\includegraphics[scale=0.5]{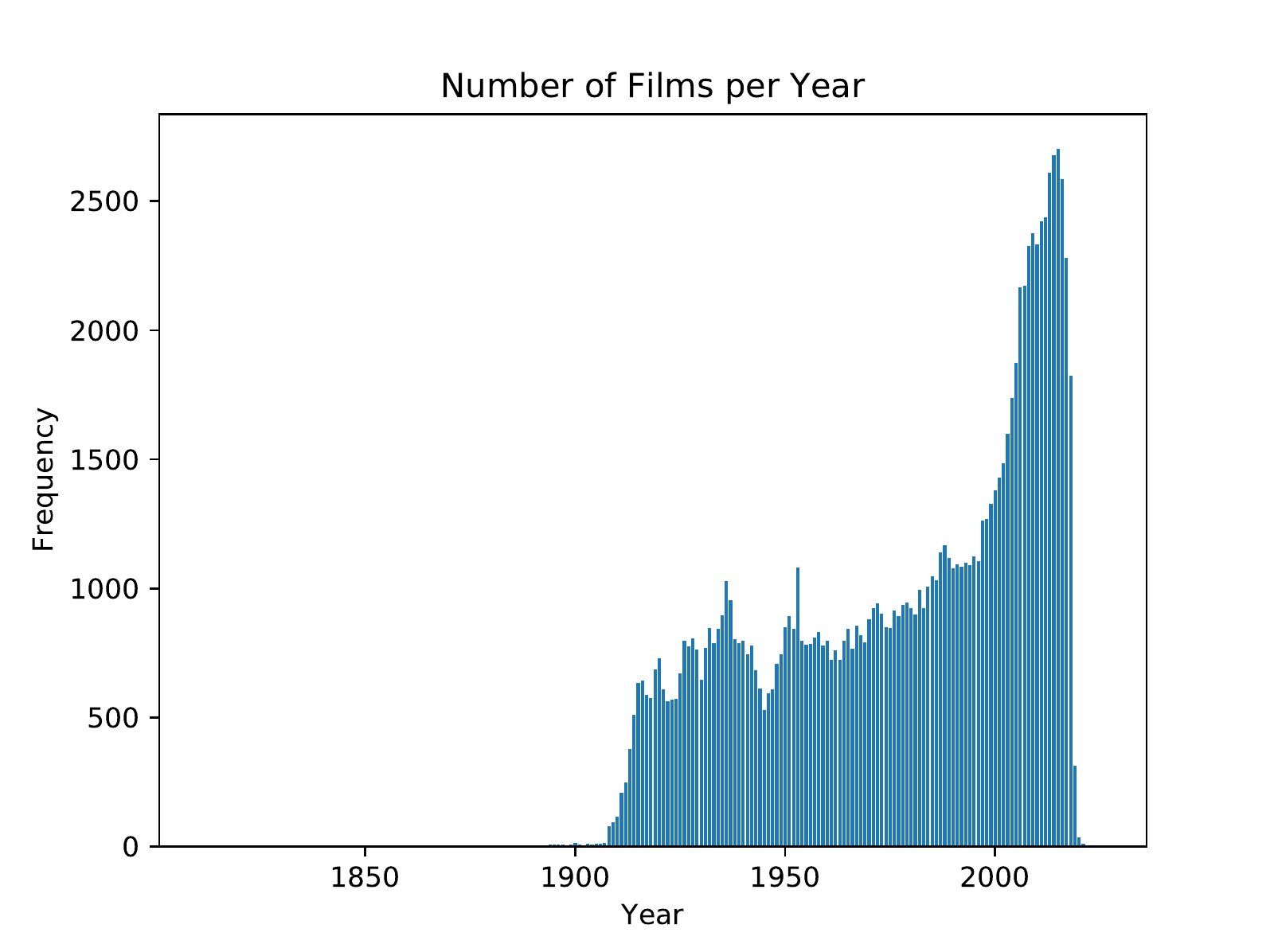} 

We now take a look at the sizes of casts used in movies. The following code produces a bar chart in the same way as the previous example. 
\begin{lstlisting}[language = python, frame=single]
C = collections.Counter([len(d["cast"]) for d in Movies])
plt.xlabel("Cast Size")
plt.ylabel("Frequency")
plt.title("Number of Movies per Cast Size")
plt.bar(list(C.keys()), list(C.values()))
plt.show()
\end{lstlisting}
This leads to the following bar chart:

\includegraphics[scale=0.5]{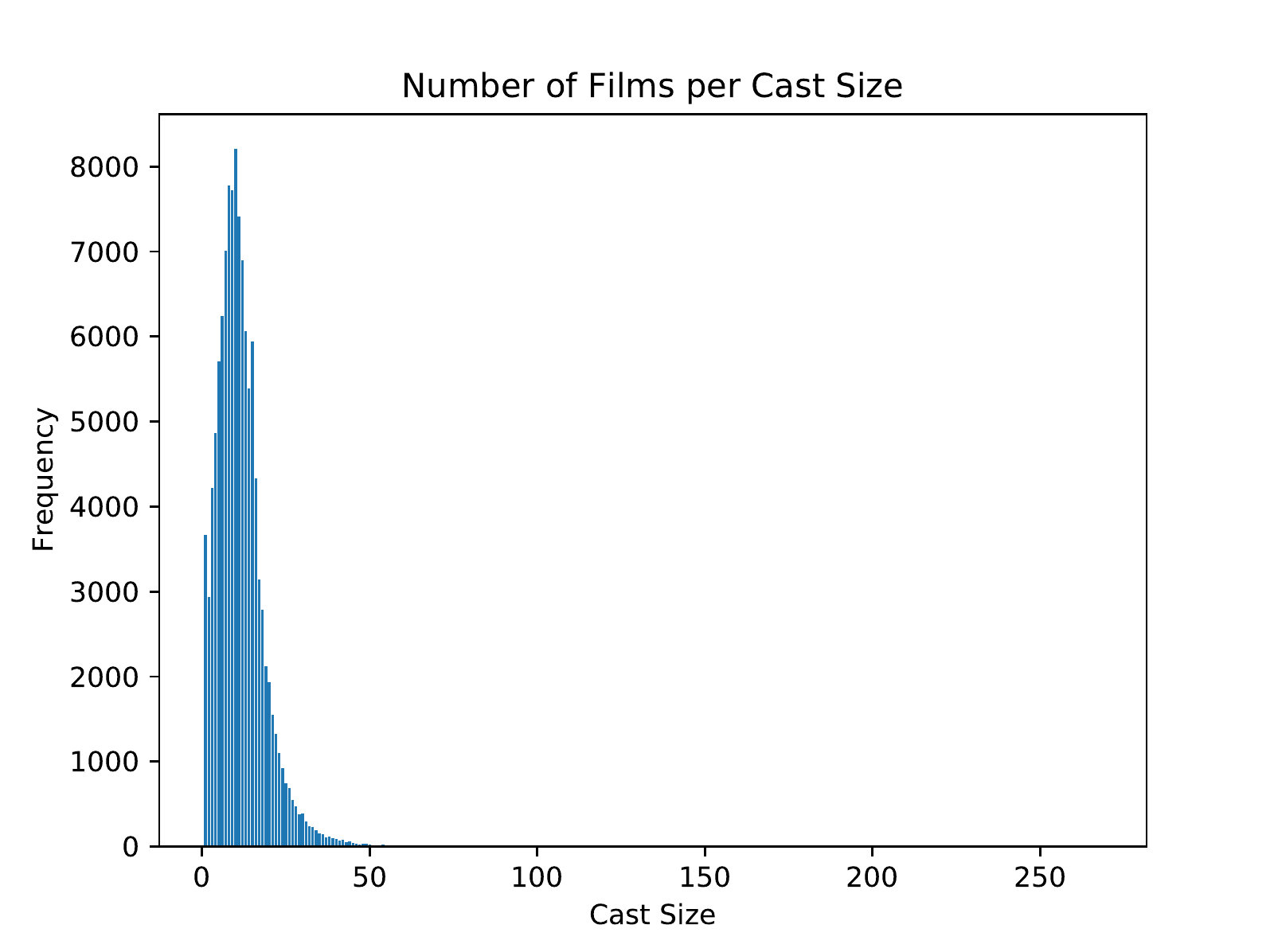} 

This indicates that nearly all movies have casts of between one and fifty actors. However, there are again some outliers with much larger casts. To get the names of these movies, the following code reorders the list \py{Movies} into descending order of cast size. The first five movies on this list are then written to the screen. 

\begin{lstlisting}[language = python, frame=single]
Movies = sorted(Movies, key=lambda i: len(i["cast"]), reverse=True)
for i in range(5):
    print(Movies[i]["title"], "=", len(Movies[i]["cast"]))
\end{lstlisting}
This produces the following output, indicating the five movies with the largest cast sizes.
\begin{lstlisting}[frame=single]
Cirque du Soleil: Worlds Away = 268
Hollywood Without Make-Up = 132
The Longest Day (film) = 117
The Founding of a Party = 116
The Founding of a Republic = 106
\end{lstlisting}

As a final point, we can also see in the above bar chart that there is a preponderance of movies with a cast size of one. In some cases this is correct, such as with the 2018 stand-up comedy movie \emph{Russell Brand: Re:Birth}. On the other hand, this also reveals some further problems in the dataset. For example, the movie \emph{Lady with a Sword} (1971) is also recorded as having a cast size of one despite the fact that many actors actually appeared in it, such as Lily Ho, James Nam and Hsieh Wang. 

%---------------------------------------------------------------------------
\section{Forming the Social Network}

In this section we now construct the complete social network of actors using our dataset together with tools available in the Python library NetworkX. As mentioned earlier, our network is made up of nodes (actors in this case), with edges connecting actors that have appeared in a movie together. Probably the most appropriate type of network to use here is a \emph{multigraph}. Multigraphs allow us to define multiple edges between the same pair of nodes, which makes sense here because actors will often appear in multiple movies together. Note also that the edges in this network are not directed. This means that if actor A has appeared with actor B, then B has also appeared with A.

The following code constructs our network \py{G} using the \py{Movies} list from the previous section. As shown, the code considers each movie in turn. It then goes through each pair of actors that appeared in this movie and adds the appropriate edge to the network. Each edge is also labelled with the corresponding movie title. Upon construction of the network, the methods \py{G.number_of_nodes()} and \py{G.number_of_edges()} are then used to output some information to the screen.

\begin{lstlisting}[language = python, frame=single]
G = nx.MultiGraph()        
for movie in Movies:
    for i in range(0, len(movie["cast"]) - 1):
        for j in range(i + 1, len(movie["cast"])):
            G.add_edge(movie["cast"][i], movie["cast"][j], title=movie["title"])

print("Number of nodes in this multigraph =", G.number_of_nodes())
print("Number of edges in this multigraph =", G.number_of_edges())
\end{lstlisting}

As shown in the following output, the resultant network is very large, with a total of 395,414 different nodes (actors) and 9,968,607 different edges.
\begin{lstlisting}[frame=single]
Number of nodes in this multigraph = 395414
Number of edges in this multigraph = 9968607
\end{lstlisting}

%---------------------------------------------------------------------------
\section{Analysing Connections in the Network}

Having formed our social network of actors, we can now analyse some of its interesting features. In this section we start by calculating the total number of movies that each actor has appeared in. We then determine the most prolific acting partnerships in the movie business by calculating the number of movies that each pair of actors has starred in.

\subsection{Movies Per Actor}

The following piece of code calculates the total number of movies per actor and lists the top five. For each node in our network, this is achieved by going through its incident edges and forming a set \py{S} of all the different labels (movie titles) appearing on these edges. The final results are stored in the dictionary \py{D}. For output purposes, the contents of \py{D} are then put into a sorted list \py{L}, and the first five entries in this list are written to the screen.

\begin{lstlisting}[language = python, frame=single]
D = {}
for v in G.nodes():
    E = list(G.edges(v, data=True))
    S = set()
    for e in E:
        S.add(e[2]["title"])
    D[v] = S
L = sorted(D.items(), key=lambda item: len(item[1]), reverse=True)
for i in range(5):
    print(L[i][0], ":", len(L[i][1]))    
\end{lstlisting}

The output below shows the results. We see that the top positions are occupied by actors from Indian cinema, with the great Sukumari (1940--2013) winning the competition with 703 recorded movie appearances. The top one-hundred actors from this list are shown in Appendix~\ref{app:movPerActor} at the end of this document. 

\begin{lstlisting}[frame=single]
Sukumari : 703
Jagathy Sreekumar : 695
Adoor Bhasi : 579
Brahmanandam : 576
Manorama : 558
\end{lstlisting}

\subsection{Acting Partnerships}

We now consider the number of collaborations between different pairs of actors---that is, the number of movies that each pair of actors has appeared in together. 

The following code calculates these figures. It goes through every pair of actors that are known to have appeared in at least one movie together, and then counts the total number of edges between the corresponding nodes. This information is collected in the dictionary \py{D}, which is again copied into an ordered list \py{L}. Again, the top five collaborations are then reported. 

\begin{lstlisting}[language = python, frame=single]
D = {}
for e in G.edges():
    D[e[0] + " and " + e[1]] = G.number_of_edges(e[0], e[1])
L = sorted(D.items(), key=lambda kv: kv[1], reverse=True)
for i in range(5):
    print(L[i][0], ":", L[i][1])
\end{lstlisting}

The output from this code is below. We see that the most prolific acting partnership in this network is due to the late Indian actors Adoor Bhasi (1927--1990) and Prem Nazir (1926--1991), who appeared in an impressive 292 movies together. Next on the list are Larry Fine and Moe Howard (two of the Three Stooges) who co-starred in 216 movies. By comparison, the comedy partnership of Oliver Hardy and Stan Laurel resulted in a paltry 105 movies, putting them at position 46 in the list overall. The top one-hundred acting partnerships are also listed in Appendix~\ref{app:movPerActor}.

\begin{lstlisting}[frame=single]
Adoor Bhasi and Prem Nazir : 292
Larry Fine and Moe Howard : 216
Adoor Bhasi and Sankaradi : 207
Adoor Bhasi and Bahadoor : 198
Brahmanandam and Ali : 193
\end{lstlisting}

%---------------------------------------------------------------------------
\section{Calculating Shortest Paths}

As we have seen, when two actors have not appeared in a movie together there will be no edge between the corresponding nodes in the social network. However, we can still look for connections between actors by using \emph{paths} of intermediate actors. This is similar to the so-called ``Six Degrees of Separation''---the idea that all people are six or fewer social connections away from each other~\cite{wiki6}.

Connecting actors using chains of intermediate actors is an idea popularised by the Oracle of Bacon website~\cite{OoB}, who provide a simple tool for finding shortest paths between any pair of actors. As mentioned earlier, the Oracle of Bacon is also the source of the dataset used in this work.

As an example, according to our dataset we find that the actors Anthony Hopkins and Samuel L. Jackson have never appeared in a movie together. In our network, this means that the corresponding two nodes have no edge between them. However, these nodes can still be regarded as fairly ``close'' to one another because, in this case, they are both linked to the node representing Scarlett Johansson. (Specifically, Anthony Hopkins acted with Scarlett Johansson in \emph{Hitchcock}, and Samuel L. Jackson appeared with Johansson in \emph{Captain America: The Winter Soldier}). The shortest path from Anthony Hopkins to Samuel L. Jackson therefore has a length of two, since we need to travel along two edges in the network to get from one actor to the other. In reality, there may be many paths between Anthony Hopkins and Samuel L. Jackson in our network. However, determining the \emph{shortest} path tells us that there are no paths with fewer edges. 

Before looking at the techniques used in identifying shortest paths, we will first simplify our network slightly by converting it into a ``simple graph''. Simple graphs allow a maximum of one edge between a pair of nodes; hence, when we have multiple edges between a pair of nodes in our multigraph (because the two actors have appeared in multiple movies together), these will now be represented as a single edge. Note that this conversion maintains the number of nodes in the network but it reduces the number of edges. It will therefore make some of our calculations a little quicker. The following code constructs our simple graph. The final line then checks whether this new network is connected. (A network is connected when it is possible to form a path between every pair of nodes.) 

\begin{lstlisting}[language = python, frame=single]
G = nx.Graph()        
for movie in Movies:
    for i in range(0, len(movie["cast"]) - 1):
        for j in range(i + 1, len(movie["cast"])):
            G.add_edge(movie["cast"][i], movie["cast"][j], title=movie["title"])         
print("Number of nodes in simple graph  =", G.number_of_nodes())
print("Number of edges in simple graph  =", G.number_of_edges())
print("Graph Connected?                 =", nx.is_connected(G))
\end{lstlisting}
This produces the following output. As can be seen, the network \py{G} still has 395,414 nodes, but it now contains 8,676,962 edges instead of the original 9,968,607---a 13\% reduction. We also see that the network is not connected; that is, it is composed of more than one distinct connected component. 

\begin{lstlisting}[frame=single]
Number of nodes in simple graph  = 395414
Number of edges in simple graph  = 8676962
Graph Connected?                 = False
\end{lstlisting}

Shortest paths can now be found in our social network using the NetworkX command \py{nx.shortest_path()}. If the edges of the graph are unweighted (as is the case here) then this invokes a breadth first search; otherwise the slightly more expensive Dijkstra's algorithm is used. Both of these methods are reviewed by Rosen~\cite{Rosen2018}. In either case, the output from this command is a list of nodes \py{P} that specifies the shortest path between the two specified nodes. For example, the code:

\begin{lstlisting}[language = python, frame=single]
P = nx.shortest_path(G, source="Anthony Hopkins", target="Samuel L. Jackson")
print(P)
\end{lstlisting}
gives the following output. 
\begin{lstlisting}[frame=single]
[Anthony Hopkins, Scarlett Johansson, Samuel L. Jackson]
\end{lstlisting}

While this code does indeed tell us the shortest path between Anthony Hopkins and Samuel L. Jackson, it does not give the names of the movies involved in this path. In addition, if no path exists between the actors, or if we type in a name that is not present in the network, then the program will halt with an exception error. A better alternative is to therefore put the \py{nx.shortest_path} statement into a bespoke function, and then add some code that (a) checks for errors, and (b) writes the output in a more helpful way. The following code does this.

\begin{lstlisting}[language = python, frame=single]
def writePath(G, u, v):
    print("Here is the shortest path from", u, "to", v, ":")
    if not u in G or not v in G:
        print("  Error:", u, "and/or", v, "are not in the network")
        return
    try:
        P = nx.shortest_path(G, source=u, target=v)
        for i in range(len(P) - 1):
            t = G.edges[P[i],P[i+1]]["title"]
            print(" ", P[i], "was in", t, "with", P[i+1])
    except nx.NetworkXNoPath:
        print("  No path exists between", u, "and", v)

writePath(G, "Catherine Zeta-Jones", "Jonathan Pryce")
writePath(G, "Homer Simpson", "Neil Armstrong")
\end{lstlisting}

The bottom two lines of the above code make two calls to the \py{writePath()} function, resulting in the following output. 
\begin{lstlisting}[frame=single]
Here is the shortest path from Catherine Zeta-Jones to Jonathan Pryce
  Catherine Zeta-Jones was in Ocean's Twelve with Albert Finney
  Albert Finney was in Loophole with Jonathan Pryce
Here is the shortest path from Homer Simpson to Neil Armstrong
  Error: Homer Simpson and/or Neil Armstrong are not in the network
\end{lstlisting}

%---------------------------------------------------------------------------
\section{Connectivity and Centrality Analysis}

In this section we now use three techniques from the field of centrality analysis to help us identify who the most ``central'' and ``important'' actors are in our social network.

Recall from the last section that our network is currently not connected. This means that the graph is made up more than one connected component, and that paths between actors in different components do not exist. To investigate these connected components, we can use the command \py{nx.connected_components(G)} to construct a list holding the number of nodes in each. Details of these can then be written to the screen:

\begin{lstlisting}[language = python, frame=single]
C = [len(c) for c in sorted(nx.connected_components(G), key=len, reverse=True)]
print("Number of components =", len(C))
print("Component sizes      =", C)
\end{lstlisting}

The output of these statements is shown below. We see that our network of actors is actually made up of 2,533 different components; however, the vast majority of the nodes (96\%) all occur within the same single connected component, indicating the existence of paths between all of these 379,859 different actors. We also see that the remaining components are very small (in most cases they are composed of actors who all appeared in the same single together movie, but no others).  

\begin{lstlisting}[frame=single]
Number of components = 2533
Component sizes      = [379859, 72, 46, 45, 42, 40,..., 2, 2]
\end{lstlisting}

For the remainder of our analysis we will now focus on the single connected component of 379,859 actors. To do this we first need to isolate this component. We can then use the \py{subgraph()} command to form the network represented by the component.

\begin{lstlisting}[language = python, frame=single]
C = max(nx.connected_components(G), key=len)
G = G.subgraph(C)
print("Number of nodes in simple graph =", G.number_of_nodes())
print("Number of edges in simple graph =", G.number_of_edges())
print("Graph Connected?                =", nx.is_connected(G))
\end{lstlisting}

The output of the above code confirms that this new network is indeed connected as we would expect.

\begin{lstlisting}[frame=single]
Number of nodes in simple graph = 379859
Number of edges in simple graph = 8612493
Graph Connected?                = True
\end{lstlisting}

The following three subsections will now investigate the ``centrality'' of the nodes appearing in this connected network. As mentioned, three different measures will be considered: degree centrality, betweenness centrality, and closeness centrality.

\subsection{Degree Centrality}

In networks, the \emph{degree} of a node is simply the number of edges that are touching it. For the network of actors that we have now formed, the degree therefore represents the number of different people that this actor has worked with. To access the degree of a node, the simplest option is to use the command \py{G.degree(v)}. For example, the code:
\begin{lstlisting}[language = python, frame=single]
print(G.degree("Henry Fonda"))
\end{lstlisting}
produces the output
\begin{lstlisting}[frame=single]
1325
\end{lstlisting}
telling us that the actor Henry Fonda has appeared in film with 1,325 different actors. A second option is to use the NetworkX function \py{nx.degree_centrality(G)}, which calculates the \emph{degree centrality} of each node in the network. The degree centrality of a node is determined by dividing the degree of the node by the maximum possible degree of a node, which in this case is simply the number of nodes in the network minus one (i.e., 379,858). The degree centrality of Henry Fonda, for example, is therefore calculated as 1,325 divided by 379,858, giving 0.00349. This figure can be interpreted as the proportion of actors in the network that Henry Fonda has acted with---in this case, just under 0.35\%. 

The following code creates a dictionary \py{D} that holds the degree centrality of all nodes in the network. As with previous examples, the top five actors according to this measure are then written to the screen. 

\begin{lstlisting}[language = python, frame=single]
D = nx.degree_centrality(G)
L = sorted(D.items(), key=lambda item: item[1], reverse=True)
for i in range(5):
    print(L[i][0], ":", L[i][1])
\end{lstlisting}

This code produces the output below. To determine the actual degree of these nodes (i.e., the number of different actors that each actor has worked with), we simply need to multiply these figures by the number of nodes minus one. We then find that Nassar has appeared with a massive 2,937 different actors; Sukumari with 2,549; Manorama with 2,511; Brahmanandam with 2,460; and Vijayakumar with 2,369. A listing of the top one-hundred actors is given in Appendix~\ref{app:CentAnal}.

\begin{lstlisting}[frame=single]
Nassar : 0.007731836633689431
Sukumari : 0.006710402308230971
Manorama : 0.006610364925840709
Brahmanandam : 0.0064761042284222
Vijayakumar : 0.006236541023224468
\end{lstlisting}

One other notable actor on this list is the English actor Christopher Lee (1922--2015), who appears at position 11, having appeared on screen with 2,056 different actors. Our reasons for mentioning Christopher Lee in particular will become apparent in the next two subsections.

\subsection{Betweenness Centrality}

Betweenness centrality considers the number of shortest paths in a network that pass through a particular node. In social networks this helps to detect the ``middlemen'' that serve as a links between different parts of a network. It also helps to identify ``hubs'' in the network that, when removed, will start to disconnect parts of the network from each other. A useful analogy can be drawn with road networks of cities. Shortest paths that travel across the city will often pass through the same locations in the road network (consider the \emph{Arc de Triomphe} in Paris, for example). As a result, the nodes at these locations can be considered more ``central'' to the network.

The formula for calculating the betweenness centrality of an individual node $v$ in a network is as follows: 
$$
C(v)=\sum_{s,t\in V}\frac{\sigma(s,t|v)}{\sigma(s,t)}
$$
where $V$ is the set of nodes in the network, $\sigma(s,t)$ is the number of different shortest paths between two nodes $s$ and $t$, and $\sigma(s,t|v)$ is the number of these $s$-$t$-paths that are seen to pass through $v$. In other words, the betweenness centrality of a node $v$ is the sum of the fraction of all-pairs shortest paths that pass through node $v$.

We can calculate the betweenness centrality of all actors in our social network using the NetworkX command \py{nx.betweenness_centrality(G)}. This uses an algorithm designed by Brandes~\cite{Brandes2001}. However, executing this command on a network as big as ours is infeasible. This is because it involves having to calculate \emph{all} of the shortest paths between \emph{all} pairs of nodes, which would take a huge amount of calculation. (In technical terms, it involves using an algorithm that has a complexity of $O(nm)$, where $n$ is the number of nodes and $m$ is the number of edges.) Luckily, we can make some savings in these calculations by using a sample of $k$ nodes to estimate the betweenness centrality of all nodes~\cite{Brandes2007}. This is carried out by the following code using a sample of 1,000 nodes. As before, the top five results are written to the screen. Some statements are also included to allow us to measure how long the calculation takes.

\begin{lstlisting}[language = python, frame=single]
start = time.time()
D = nx.betweenness_centrality(G, k=1000)
end = time.time()
print("Time taken =", end-start, "seconds")
L = sorted(D.items(), key=lambda item: item[1], reverse=True)
for i in range(5):
    print(L[i][0], ":", L[i][1])
\end{lstlisting}

The output of this code is shown below. It indicates that Christopher Lee appears on by far the highest number of shortest paths in the network, followed by Om Puri, Jackie Chan, Anupam Kher, and then Harrison Ford. As indicated, this calculation took approximately 10 hours to carry out on our computer (a 3.2 GHtz Windows 10 machines with 8 GB RAM). This suggests that a full calculation using all nodes instead of just a sample would have taken something in the region of 150 days to complete. The top one-hundred actors for this measure are listed in Appendix~\ref{app:CentAnal}.

\begin{lstlisting}[frame=single]
Time taken = 35078.12739729881 seconds
Christopher Lee : 0.009681571845060294
Om Puri : 0.008096614699536075
Jackie Chan : 0.007916442261041035
Anupam Kher : 0.007905221698069204
Harrison Ford : 0.004888140405090745
\end{lstlisting}

\subsection{Closeness Centrality}

Like betweenness centrality, closeness centrality also considers shortest paths in a network. For a given node $v$ it represents the mean shortest path length from $v$ to all other nodes in the network~\cite{Freeman1979,Wasserman1994}. If an actor is found to be connected to other actors via  short paths, they can therefore be considered to be quite central in the network. The formula used for calculating the closeness centrality of a particular node $v$ is as follows:
$$
C(v) = \frac{1}{\left(\sum_{u\in V}d(v,u)\right)/n}
$$
where $d(v,u)$ is the length of the shortest path (number of edges) between nodes $v$ and $u$, and $n$ is the number of nodes in the network. Higher values of this measure therefore indicate a higher centrality. Note that we can also calculate the actual mean path length from a node $v$ to all other nodes in the network by simply dividing 1 by $C(v)$. 

As with the previous example, calculating the closeness centrality of all nodes in our large network of actors would take too long on a single computer because, once again, it involves calculating the shortest paths between all pairs of nodes. In our case we make things easier by restricting our calculations to the top 1,000 actors according to the betweenness centrality measure from the previous subsection. The following code does this. First, it produces a list \py{V} of all actors in the network, ordered according to their betweenness centrality score. The closeness centrality is then calculated for each of the first 1,000 actors in this list. These are then ranked, and the top five are output. 

\begin{lstlisting}[language = python, frame=single]
V = [L[i][0] for i in range(len(L))]
D = {}
start = time.time()
for i in range(1000):
    D[V[i]] = nx.closeness_centrality(G, V[i])
end = time.time()
print("Time taken =", end-start, "seconds")
L = sorted(D.items(), key=lambda item: item[1], reverse=True)
for i in range(5):
    print(L[i][0], ":", L[i][1])
\end{lstlisting}

The following output shows that, of these actors, Christopher Lee is again the most central. Amazingly, we find that we can get from Christopher Lee to any other actor in the network in an average of just 2.88 hops. The next best-connected actors are then, respectively, Michael Caine (average of 2.917 hops), Harvey Keitel (2.922), Christopher Plummer (2.931) and Robert De~Niro (2.936). The top one-hundred actors according to this measure are also listed in Appendix~\ref{app:CentAnal}.

\begin{lstlisting}[frame=single]
Time taken = 29075.51345229149 seconds
Christopher Lee : 0.34724000968978963
Michael Caine : 0.3428649311215694
Harvey Keitel : 0.3422440138966974
Christopher Plummer : 0.34123099284135183
Robert De Niro : 0.34059459561239835
\end{lstlisting}

If we want, we can also take a closer look at the number of actors within a certain number of hops from a chosen actor. For example, the following code creates a dictionary \py{D} that holds the length of the shortest path between Christopher Lee and all other actors in the network. It then counts the number of actors of distance 0, 1, 2, and so on.

\begin{lstlisting}[language = python, frame=single]
D = nx.shortest_path_length(G, "Christopher Lee")
print(collections.Counter(D.values()))
\end{lstlisting}

The output from this code tells us that all actors can be reached from Christopher Lee using fewer than nine hops. Exactly one actor can be reached in zero hops (Christopher Lee himself); 2,056 actors can be reached with one hop; 98,758 with two hops; and so on. In particular, note that over 86\% of actors can be reached from Christopher Lee in fewer than four hops, and 99\% in fewer than five hops. 

\begin{lstlisting}[frame=single]
Counter({0: 1, 1: 2056, 2: 98758, 3: 226350, 4: 48625, 5: 3655,  6: 369, 7: 36, 8: 9})
\end{lstlisting}

\subsection{Distribution of Actors' Closeness Centrality Scores}

Finally, it is also interesting to look at the distribution of different actors' closeness centrality scores. We have already seen that actors like Christopher Lee, Om Puri, and Michael Caine are very central and well connected, but what is the score of a ``typical'' actor. Once again, the expense of calculating shortest paths between all pairs of actors is prohibitively expensive for a large network like ours. Instead, the code below takes a random sample of 1,000 actors, calculates their closeness centrality scores, works out the mean and standard deviation of this sample, and then uses the bespoke function \py{doHistogramSummary()} to plot this information to the screen.

\begin{lstlisting}[language = python, frame=single]
def doHistogramSummary(X, xlabel, ylabel, title):
    plt.xlabel(xlabel)
    plt.ylabel(ylabel)
    plt.title(title)
    plt.hist(X, bins=20)
    plt.show()

random.shuffle(V)
D = {}
for i in range(1000):
    D[V[i]] = 1 / nx.closeness_centrality(G, V[i])
L = D.values()
print("Mean =", statistics.mean(L))
print("SD   =", statistics.stdev(L))
doHistogramSummary(L, "Average distance to all Actors", "Frequency", "Closeness centrality distribution")
\end{lstlisting}

The output from this code is shown below. As we can see, the average distance between any two actors in this sample is just over 4.27 hops. As a comparison, this is slightly lower than the six hops hypothesised in the Six Degrees of Separation; however, it is slightly higher than the mean of 3.57 found in a similar analysis of Facebook friendships carried out by Facebook Research in 2016~\cite{facebook}.

\begin{lstlisting}[frame=single]
Mean = 4.2704787482812225
SD   = 0.4701311459607602
\end{lstlisting}

\includegraphics[scale=0.5]{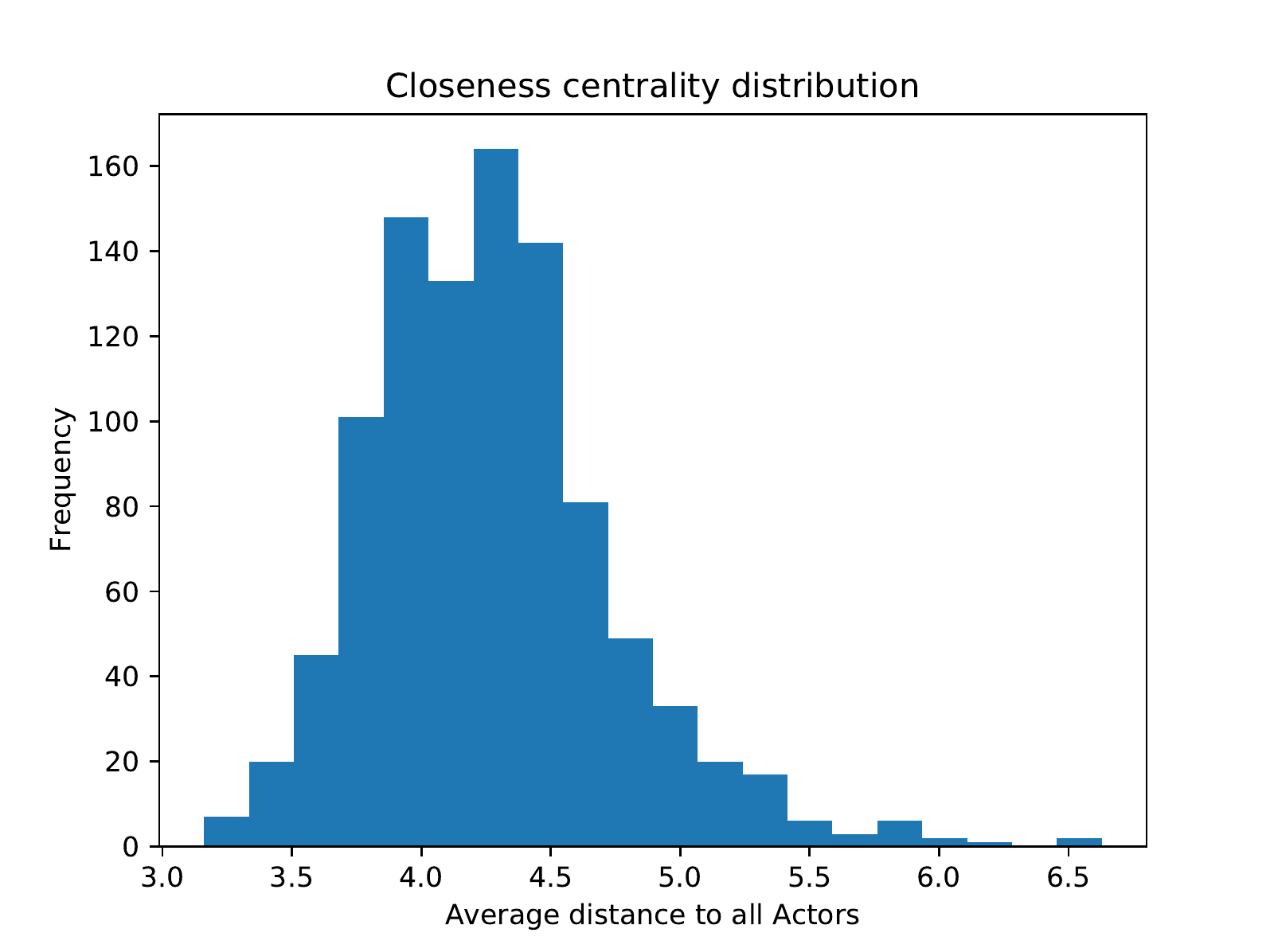}

%---------------------------------------------------------------------------
\section{Conclusions and Further Discussion}

This article has used tools from the NetworkX library to help determine the most important people in the social network of movie actors. Regardless of the measures used, the most central actors are consistently those who have or had very long acting careers, such as Christopher Lee, Nassar, Sukumari, Michael Caine, Om Puri, and Jackie Chan. This is quite natural, because long careers bring more acting opportunities, helping to improve an actor's connectivity in the network. To contrast these findings, Appendix~\ref{app:CentAnalDecade} shows the twenty most central actors by decade. These statistics were found in the same way as above, but only used movies that were released in that particular decade. As we might expect, this causes many new names to crop up. For movies released in the 1950's, for example, actors such as Louis de~Funes (1914--1983) and George Thorpe (1891--1961) seem to be very central; for the 1990's on the other hand, the most central actors are people like Samuel L. Jackson, Om Puri, Vijayakumar, Roshan Seth and Frank Welker.

There are other ways in which we might have performed this analysis. Alternative centrality measures are also included in the NetworkX library, such as page rank centrality, Eigenvector centrality, Katz centrality, and current-flow closeness centrality. A good review of these measures can be found in the book of Needham and Hodler~\cite{Needham2019}. In the future, it would also be interesting to add some kind of values to the edges of the network in order to give more information about the nature of an acting collaboration. This could include the number of minutes that both actors appeared on screen, the critical ratings of the movie, or the financial earnings. Such factors would certainly influence the set of central actors that are identified.

All materials from this study are available online:
\begin{itemize}
\item A complete listing of the dataset, code, and results tables can be found at \url{http://www.rhydlewis.eu/movies/all.zip},
\item A shorter web version of this document can be found at \url{http://www.rhydlewis.eu/movies}.
\end{itemize}

%%%%%%%%%%%%%%%%%%%%%%%%%%%%%%%%%%%%%%%%%%%%%%%%%%%%%%%%%%%%%%%%%%%%%%%%%%%%%%%%%
\bibliography{refs}
\bibliographystyle{plain}

%%%%%%%%%%%%%%%%%%%%%%%%%%%%%%%%%%%%%%%%%%%%%%%%%%%%%%%%%%%%%%%%%%%%%%%%%%%%%%%%%
\appendix

\footnotesize

\section{Movies per Actor and per Partnerships}
\label{app:movPerActor}

The following table shows the top one hundred actors (left) and acting partnerships (right) according to the number of movies they have appeared in. Note that position 68 in the right list is occupied by a pair of actors called ``TBA and TBA''. This is clearly a fault in the underlying dataset.
\begin{longtable}{|l|ll|ll|} 
\hline\hline
\multicolumn{1}{|c|}{\#} &
\multicolumn{2}{c|}{Movies Per Actor} &
\multicolumn{2}{c|}{Movies Per Acting Partnership}\\ 
\hline
1	&	Sukumari	&	703	&	Adoor Bhasi and Prem Nazir	&	292	\\
2	&	Jagathy Sreekumar	&	695	&	Larry Fine and Moe Howard	&	216	\\
3	&	Adoor Bhasi	&	579	&	Adoor Bhasi and Sankaradi	&	207	\\
4	&	Brahmanandam	&	576	&	Adoor Bhasi and Bahadoor	&	198	\\
5	&	Manorama	&	558	&	Brahmanandam and Ali	&	193	\\
6	&	Sankaradi	&	545	&	Brahmanandam and Kota Srinivasa Rao	&	170	\\
7	&	Prem Nazir	&	518	&	Adoor Bhasi and K. P. Ummer	&	170	\\
8	&	Nedumudi Venu	&	470	&	Prem Nazir and Bahadoor	&	163	\\
9	&	Bahadoor	&	450	&	Brahmanandam and Tanikella Bharani	&	157	\\
10	&	Nassar	&	438	&	Adoor Bhasi and Jayabharathi	&	156	\\
11	&	Meena	&	413	&	Sankaradi and Bahadoor	&	155	\\
12	&	Mammootty	&	395	&	Senthil and Goundamani	&	151	\\
13	&	Senthil	&	381	&	Harold Lloyd and Snub Pollard	&	148	\\
14	&	Nagesh	&	375	&	Prem Nazir and Sankaradi	&	147	\\
15	&	Oliver Hardy	&	373	&	Bebe Daniels and Harold Lloyd	&	146	\\
16	&	Innocent	&	371	&	Bebe Daniels and Snub Pollard	&	146	\\
17	&	Vijayakumar	&	368	&	Sukumari and Sankaradi	&	145	\\
18	&	Shakti Kapoor	&	365	&	K. P. Ummer and Prem Nazir	&	144	\\
19	&	Madhu	&	357	&	Jagathy Sreekumar and Sukumari	&	141	\\
20	&	Mohanlal	&	348	&	Meena and Adoor Bhasi	&	139	\\
21	&	Mithun Chakraborty	&	347	&	Jayabharathi and Prem Nazir	&	138	\\
22	&	Kota Srinivasa Rao	&	346	&	Sukumari and Adoor Bhasi	&	134	\\
23	&	Ali	&	343	&	Sheela and Adoor Bhasi	&	127	\\
24	&	Srividya	&	333	&	Jayabharathi and Bahadoor	&	126	\\
25	&	Prakash Raj	&	329	&	Adoor Bhasi and Sreelatha Namboothiri	&	123	\\
26	&	Kuthiravattam Pappu	&	329	&	Sheela and Prem Nazir	&	122	\\
27	&	Tanikella Bharani	&	324	&	Harold Lloyd and Sammy Brooks	&	120	\\
28	&	Thilakan	&	319	&	Brahmanandam and M. S. Narayana	&	120	\\
29	&	Jayabharathi	&	318	&	Meena and Prem Nazir	&	120	\\
30	&	K. P. Ummer	&	316	&	Meena and Sankaradi	&	119	\\
31	&	Mala Aravindan	&	312	&	Snub Pollard and Sammy Brooks	&	115	\\
32	&	Kaviyoor Ponnamma	&	310	&	Moe Howard and Curly Howard	&	114	\\
33	&	Siddique	&	301	&	Bebe Daniels and Sammy Brooks	&	113	\\
34	&	Rekha	&	296	&	Larry Fine and Curly Howard	&	113	\\
35	&	Sivaji Ganesan	&	295	&	Mohanlal and Sukumari	&	110	\\
36	&	Anupam Kher	&	290	&	Jagathy Sreekumar and Innocent	&	110	\\
37	&	Lakshmi	&	288	&	Jayabharathi and Sankaradi	&	110	\\
38	&	N. T. Rama Rao	&	285	&	Roscoe 'Fatty' Arbuckle and Fatty Arbuckle	&	109	\\
39	&	Indrans	&	283	&	Adoor Bhasi and Thikkurissi Sukumaran Nair	&	109	\\
40	&	M. N. Nambiar	&	280	&	Brahmanandam and Venu Madhav	&	108	\\
41	&	Ambika	&	276	&	Meena and Bahadoor	&	108	\\
42	&	Jose Prakash	&	276	&	Paravoor Bharathan and Sankaradi	&	108	\\
43	&	Prathapachandran	&	276	&	Jagathy Sreekumar and Indrans	&	107	\\
44	&	Asrani	&	275	&	Prem Nazir and Jose Prakash	&	107	\\
45	&	MG Soman	&	272	&	Sukumari and Nedumudi Venu	&	106	\\
46	&	Vadivelu	&	271	&	Oliver Hardy and Stan Laurel	&	105	\\
47	&	Vijayaraghavan	&	270	&	Jagathy Sreekumar and Nedumudi Venu	&	105	\\
48	&	Sheela	&	265	&	Madhu and Adoor Bhasi	&	104	\\
49	&	Saikumar	&	264	&	Adoor Bhasi and Paravoor Bharathan	&	104	\\
50	&	Aruna Irani	&	264	&	Thikkurissi Sukumaran Nair and Prem Nazir	&	104	\\
51	&	Mukesh	&	261	&	Madhu and Sankaradi	&	103	\\
52	&	Devan	&	258	&	K. P. Ummer and Bahadoor	&	103	\\
53	&	Paravoor Bharathan	&	257	&	Prem Nazir and T. S. Muthaiah	&	101	\\
54	&	Venniradai Moorthy	&	256	&	Ali and Kota Srinivasa Rao	&	100	\\
55	&	Rajkumar	&	256	&	Brahmanandam and Raghu Babu	&	99	\\
56	&	Gulshan Grover	&	256	&	Kaviyoor Ponnamma and Sankaradi	&	99	\\
57	&	Goundamani	&	254	&	Innocent and Sukumari	&	98	\\
58	&	Delhi Ganesh	&	252	&	Sreelatha Namboothiri and Prem Nazir	&	98	\\
59	&	V. K. Ramasamy	&	252	&	Salim Kumar and Jagathy Sreekumar	&	97	\\
60	&	Kader Khan	&	251	&	Mammootty and Sukumari	&	97	\\
61	&	Murali	&	247	&	Ali and Tanikella Bharani	&	97	\\
62	&	Raymond Hatton	&	247	&	N. T. Rama Rao and Relangi	&	97	\\
63	&	Pandari Bai	&	246	&	Innocent and Nedumudi Venu	&	96	\\
64	&	Dharmendra	&	244	&	Kaviyoor Ponnamma and Adoor Bhasi	&	96	\\
65	&	Kovai Sarala	&	244	&	Adoor Bhasi and Jose Prakash	&	96	\\
66	&	Harry Carey	&	244	&	K. P. Ummer and Sankaradi	&	96	\\
67	&	KPAC Lalitha	&	244	&	Kader Khan and Shakti Kapoor	&	96	\\
68	&	Salim Kumar	&	242	&	TBA and TBA	&	95	\\
69	&	Manobala	&	242	&	Sankaradi and MG Soman	&	94	\\
70	&	Kalpana	&	241	&	Adoor Bhasi and T. R. Omana	&	92	\\
71	&	Balakrishna	&	241	&	K. S. Ashwath and Rajkumar	&	92	\\
72	&	Mamukkoya	&	240	&	Allu Ramalingaiah and N. T. Rama Rao	&	91	\\
73	&	Prem Chopra	&	240	&	Mohanlal and Nedumudi Venu	&	90	\\
74	&	Cochin Haneefa	&	239	&	Satyanarayana and N. T. Rama Rao	&	90	\\
75	&	Avinash	&	239	&	Madhu and Bahadoor	&	89	\\
76	&	Allu Ramalingaiah	&	236	&	Adoor Bhasi and T. S. Muthaiah	&	89	\\
77	&	K. S. Ashwath	&	236	&	Harold Lloyd and Charles Stevenson	&	88	\\
78	&	Charle	&	236	&	Brahmanandam and Jaya Prakash Reddy	&	88	\\
79	&	Akkineni Nageswara Rao	&	234	&	Brahmanandam and Chalapathi Rao	&	88	\\
80	&	Manivannan	&	232	&	Sukumari and Thilakan	&	88	\\
81	&	Seema	&	232	&	Manorama and Nagesh	&	88	\\
82	&	Jagadish	&	231	&	K. P. Ummer and Jayabharathi	&	88	\\
83	&	Vivek	&	230	&	Akkineni Nageswara Rao and Gummadi	&	88	\\
84	&	Chalapathi Rao	&	230	&	M. S. Narayana and Ali	&	87	\\
85	&	M. S. Narayana	&	229	&	Sukumari and Bahadoor	&	87	\\
86	&	Jayaram	&	228	&	Satyanarayana and Allu Ramalingaiah	&	87	\\
87	&	Ashok Kumar	&	228	&	Mikkilineni and N. T. Rama Rao	&	86	\\
88	&	Sukumaran	&	227	&	Balakrishna and Rajkumar	&	86	\\
89	&	Chandra Mohan	&	226	&	Narasimharaju and Rajkumar	&	86	\\
90	&	Thikkurissi Sukumaran Nair	&	224	&	Venu Madhav and Ali	&	85	\\
91	&	Kunchan	&	224	&	Mohanlal and Jagathy Sreekumar	&	85	\\
92	&	Bebe Daniels	&	223	&	Jagathy Sreekumar and Jagadish	&	85	\\
93	&	Nizhalgal Ravi	&	223	&	Joe Cobb and Allen Hoskins	&	85	\\
94	&	Jeetendra	&	222	&	Jagathy Sreekumar and Thilakan	&	84	\\
95	&	Moe Howard	&	221	&	Adoor Bhasi and Sathyan	&	84	\\
96	&	Kalabhavan Mani	&	220	&	Brahmanandam and Dharmavarapu Subramanyam	&	83	\\
97	&	Satyanarayana	&	219	&	Kota Srinivasa Rao and Tanikella Bharani	&	83	\\
98	&	Sreelatha Namboothiri	&	219	&	Adoor Bhasi and S. P. Pillai	&	83	\\
99	&	Lalu Alex	&	218	&	Prem Nazir and S. P. Pillai	&	83	\\
100	&	Larry Fine	&	217	&	Meena and K. P. Ummer	&	82	\\
\hline
\end{longtable}

\section{Centrality Analysis of the Entire Network}
\label{app:CentAnal}

The following table lists the top one hundred actors according to degree centrality, closeness centrality and betweenness centrality. These statistics were calculated on the single connected network comprising 379,859 nodes (actors) and 8,612,493 edges. Note the presence of some spurious entries in the table such as ``Tarzan'', ``King Kong'', and ``Sam'', which reveal some problems with the underlying dataset.
\begin{longtable}{|l|ll|ll|ll|} 
\hline\hline
\multicolumn{1}{|}{} &
\multicolumn{2}{c}{Degree Centrality} &
\multicolumn{2}{c}{} &
\multicolumn{2}{c|}{Closeness Centrality} \\
\multicolumn{1}{|}{} &
\multicolumn{2}{c}{(Num. Acting Collaborations)} &
\multicolumn{2}{c}{Betweenness Centrality} &
\multicolumn{2}{c|}{(Mean Path Length)} \\
\hline
\#	& Actor & Score & Actor & Score & Actor & Score \\
\hline
1	&	Nassar	&	2937	&	Christopher Lee	&	0.00968	&	Christopher Lee		&	2.880	\\
2	&	Sukumari	&	2549	&	Om Puri	&	0.00810	&	Michael Caine		&	2.917	\\
3	&	Manorama	&	2511	&	Jackie Chan	&	0.00792	&	Harvey Keitel		&	2.922	\\
4	&	Brahmanandam	&	2460	&	Anupam Kher	&	0.00791	&	Christopher Plummer		&	2.931	\\
5	&	Vijayakumar	&	2369	&	Harrison Ford	&	0.00489	&	Robert De Niro		&	2.936	\\
6	&	Prakash Raj	&	2349	&	Klaus Kinski	&	0.00470	&	Donald Sutherland		&	2.937	\\
7	&	Jagathy Sreekumar	&	2223	&	Nassar	&	0.00468	&	Harrison Ford		&	2.943	\\
8	&	Mithun Chakraborty	&	2142	&	Tarzan	&	0.00467	&	John Hurt		&	2.954	\\
9	&	Rekha	&	2116	&	Marcello Mastroianni	&	0.00448	&	Willem Dafoe		&	2.964	\\
10	&	Nedumudi Venu	&	2063	&	Rutger Hauer	&	0.00448	&	Martin Sheen		&	2.965	\\
11	&	Christopher Lee	&	2056	&	Kabir Bedi	&	0.00437	&	Sean Connery		&	2.972	\\
12	&	John Carradine	&	2027	&	Jeanne Moreau	&	0.00405	&	Samuel L. Jackson		&	2.975	\\
13	&	Meena	&	1944	&	Gerard Depardieu	&	0.00403	&	Anthony Hopkins		&	2.976	\\
14	&	Anupam Kher	&	1943	&	Isabelle Huppert	&	0.00395	&	Ben Kingsley		&	2.977	\\
15	&	Emory Parnell	&	1912	&	George Kennedy	&	0.00394	&	Max von Sydow		&	2.979	\\
16	&	Frank Welker	&	1912	&	Stellan Skarsgard	&	0.00371	&	Ernest Borgnine		&	2.981	\\
17	&	Andy Lau	&	1908	&	Saeed Jaffrey	&	0.00371	&	Dennis Hopper		&	2.985	\\
18	&	Ali	&	1894	&	Max von Sydow	&	0.00361	&	David Carradine		&	2.988	\\
19	&	Lakshmi	&	1884	&	Ben Kingsley	&	0.00348	&	Terence Stamp		&	2.988	\\
20	&	John Wayne	&	1883	&	Shakti Kapoor	&	0.00348	&	John Gielgud		&	2.992	\\
21	&	Ashish Vidyarthi	&	1869	&	Louis de Funes	&	0.00337	&	Trevor Howard		&	3.003	\\
22	&	Tanikella Bharani	&	1844	&	Raja	&	0.00332	&	Vanessa Redgrave		&	3.003	\\
23	&	Kota Srinivasa Rao	&	1837	&	Geraldine Chaplin	&	0.00328	&	Gerard Depardieu		&	3.004	\\
24	&	Shakti Kapoor	&	1837	&	Joseph	&	0.00318	&	Stellan Skarsgard		&	3.004	\\
25	&	Saikumar	&	1833	&	Shabana Azmi	&	0.00315	&	Frank Welker		&	3.010	\\
26	&	Mammootty	&	1816	&	Donald Sutherland	&	0.00309	&	Peter O'Toole		&	3.011	\\
27	&	Devan	&	1813	&	Michael Caine	&	0.00305	&	Donald Pleasence		&	3.011	\\
28	&	Nagesh	&	1791	&	Amrish Puri	&	0.00303	&	James Caan		&	3.011	\\
29	&	Manobala	&	1776	&	Jean-Claude Van Damme	&	0.00296	&	Alfred Molina		&	3.015	\\
30	&	Kalpana	&	1774	&	Alex	&	0.00294	&	George Kennedy		&	3.016	\\
31	&	Paul Fix	&	1766	&	John Savage	&	0.00286	&	Sylvester Stallone		&	3.016	\\
32	&	Irving Bacon	&	1760	&	Michel Piccoli	&	0.00284	&	Anthony Quinn		&	3.018	\\
33	&	Delhi Ganesh	&	1757	&	I. S. Johar	&	0.00282	&	Rutger Hauer		&	3.018	\\
34	&	Mohanlal	&	1756	&	Martin Sheen	&	0.00280	&	Roger Moore		&	3.019	\\
35	&	J. Farrell MacDonald	&	1749	&	Naseeruddin Shah	&	0.00279	&	John Cleese		&	3.019	\\
36	&	Siddique	&	1740	&	Gulshan Grover	&	0.00279	&	Roddy McDowall		&	3.022	\\
37	&	Om Puri	&	1702	&	George Baker	&	0.00275	&	Ned Beatty		&	3.022	\\
38	&	Geetha	&	1678	&	Dennis Hopper	&	0.00273	&	Robert Duvall		&	3.022	\\
39	&	Indrans	&	1664	&	Harvey Keitel	&	0.00270	&	Liam Neeson		&	3.022	\\
40	&	Raymond Hatton	&	1657	&	Jason Flemyng	&	0.00266	&	Michael Madsen		&	3.023	\\
41	&	Jackie Chan	&	1652	&	Thomas Kretschmann	&	0.00262	&	Malcolm McDowell		&	3.023	\\
42	&	Eric Tsang	&	1651	&	Haluk Bilginer	&	0.00259	&	Harry Dean Stanton		&	3.023	\\
43	&	Louis de Funes	&	1646	&	Liam Neeson	&	0.00255	&	Orson Welles		&	3.025	\\
44	&	Madhu	&	1639	&	King Kong	&	0.00253	&	Elliott Gould		&	3.025	\\
45	&	Vijayaraghavan	&	1639	&	Mithun Chakraborty	&	0.00253	&	Geraldine Chaplin		&	3.026	\\
46	&	Mickey Rooney	&	1631	&	Daniel Olbrychski	&	0.00252	&	John Mills		&	3.026	\\
47	&	Charles Lane	&	1613	&	Udo Kier	&	0.00252	&	Rod Steiger		&	3.026	\\
48	&	Jackie Shroff	&	1600	&	Rade Serbedzija	&	0.00251	&	James Mason		&	3.026	\\
49	&	Ambika	&	1599	&	Charlotte Rampling	&	0.00247	&	Richard Attenborough		&	3.027	\\
50	&	Innocent	&	1596	&	Danny Trejo	&	0.00242	&	Udo Kier		&	3.028	\\
51	&	Michael Caine	&	1595	&	Ahn Sung-ki	&	0.00242	&	Christopher Lloyd		&	3.032	\\
52	&	Danny Trejo	&	1595	&	Claudia Cardinale	&	0.00242	&	Michael Hordern		&	3.032	\\
53	&	Senthil	&	1592	&	Omar Sharif	&	0.00241	&	John Carradine		&	3.032	\\
54	&	Jayaram	&	1581	&	David Carradine	&	0.00240	&	James Fox		&	3.033	\\
55	&	Gulshan Grover	&	1579	&	Sean Connery	&	0.00237	&	Eli Wallach		&	3.033	\\
56	&	Marcello Mastroianni	&	1573	&	John Carradine	&	0.00236	&	Joss Ackland		&	3.033	\\
57	&	Kovai Sarala	&	1558	&	Roshan Seth	&	0.00225	&	Michael York		&	3.034	\\
58	&	Murali	&	1557	&	Anthony Quinn	&	0.00223	&	Bruce Dern		&	3.035	\\
59	&	Pierre Watkin	&	1554	&	Prem Chopra	&	0.00222	&	Christopher Walken		&	3.036	\\
60	&	Vivek	&	1551	&	Ashish Vidyarthi	&	0.00222	&	David Niven		&	3.036	\\
61	&	Ward Bond	&	1544	&	Tom Alter	&	0.00221	&	Klaus Kinski		&	3.036	\\
62	&	Srividya	&	1539	&	Franco Nero	&	0.00220	&	Omar Sharif		&	3.036	\\
63	&	Dharmendra	&	1528	&	John Hurt	&	0.00219	&	Paul Guilfoyle		&	3.037	\\
64	&	Gerard Depardieu	&	1514	&	Shin Seong-il	&	0.00217	&	Morgan Freeman		&	3.037	\\
65	&	Samuel L. Jackson	&	1513	&	Michael Kelly	&	0.00217	&	Alan Arkin		&	3.038	\\
66	&	Avinash	&	1496	&	Shashi Kapoor	&	0.00215	&	John Cusack		&	3.038	\\
67	&	Mukesh	&	1485	&	Bruno Ganz	&	0.00213	&	Bruce Willis		&	3.040	\\
68	&	Nizhalgal Ravi	&	1484	&	Eric Roberts	&	0.00212	&	Ron Perlman		&	3.041	\\
69	&	Rajesh	&	1475	&	Mohan Agashe	&	0.00212	&	Keith David		&	3.041	\\
70	&	Sharat Saxena	&	1475	&	Maria	&	0.00212	&	Charlton Heston		&	3.042	\\
71	&	Klaus Kinski	&	1471	&	Rohini Hattangadi	&	0.00210	&	Gene Hackman		&	3.042	\\
72	&	Russell Hicks	&	1470	&	Trevor Howard	&	0.00210	&	Robert Mitchum		&	3.042	\\
73	&	Naseeruddin Shah	&	1466	&	Paul Guilfoyle	&	0.00210	&	Jason Robards		&	3.043	\\
74	&	Asrani	&	1462	&	Willem Dafoe	&	0.00207	&	Steve Buscemi		&	3.044	\\
75	&	George Chandler	&	1461	&	Christopher Plummer	&	0.00207	&	John Malkovich		&	3.044	\\
76	&	Prabhu	&	1459	&	Sam Anderson	&	0.00204	&	Jack Nicholson		&	3.047	\\
77	&	Donald Sutherland	&	1453	&	Fernando Rey	&	0.00204	&	Danny Trejo		&	3.047	\\
78	&	Hubert von Meyerinck	&	1451	&	Anthony Hopkins	&	0.00204	&	Sam Neill		&	3.047	\\
79	&	Anthony Quinn	&	1443	&	Anil Kapoor	&	0.00202	&	Helen Mirren		&	3.047	\\
80	&	Thilakan	&	1434	&	Vittorio Gassman	&	0.00201	&	Burt Reynolds		&	3.048	\\
81	&	James Mason	&	1433	&	Dharmendra	&	0.00201	&	Jack Palance		&	3.049	\\
82	&	Harold Goodwin	&	1429	&	Eric Tsang	&	0.00200	&	Antonio Banderas		&	3.049	\\
83	&	Donald Pleasence	&	1427	&	Juliette Binoche	&	0.00200	&	Ian Holm		&	3.049	\\
84	&	Sayaji Shinde	&	1426	&	Richard Attenborough	&	0.00199	&	Vernon Dobtcheff		&	3.050	\\
85	&	J. Carrol Naish	&	1425	&	Akim Tamiroff	&	0.00196	&	Claudia Cardinale		&	3.050	\\
86	&	Wade Boteler	&	1417	&	Giancarlo Giannini	&	0.00195	&	Mickey Rooney		&	3.050	\\
87	&	Prem Chopra	&	1416	&	Sam	&	0.00194	&	Denholm Elliott		&	3.051	\\
88	&	Douglas Fowley	&	1413	&	Peter Stormare	&	0.00192	&	Robert Morley		&	3.052	\\
89	&	Andy Devine	&	1409	&	Ernest Borgnine	&	0.00190	&	Derek Jacobi		&	3.052	\\
90	&	Venniradai Moorthy	&	1409	&	Mickey Rooney	&	0.00189	&	Shirley MacLaine		&	3.052	\\
91	&	Cochin Haneefa	&	1407	&	Michael Hordern	&	0.00188	&	Bruce McGill		&	3.053	\\
92	&	Chandra Mohan	&	1406	&	Samuel L. Jackson	&	0.00187	&	Nicolas Cage		&	3.054	\\
93	&	Paul Guilfoyle	&	1403	&	John Wood	&	0.00187	&	Kiefer Sutherland		&	3.054	\\
94	&	Thurston Hall	&	1400	&	Terence Stamp	&	0.00187	&	Stacy Keach		&	3.054	\\
95	&	Ashok Kumar	&	1398	&	Jean Rochefort	&	0.00186	&	Jim Broadbent		&	3.054	\\
96	&	Regis Toomey	&	1396	&	Walter Rilla	&	0.00185	&	Martin Landau		&	3.054	\\
97	&	Keith David	&	1390	&	Jean Marais	&	0.00185	&	Charlotte Rampling		&	3.056	\\
98	&	M. N. Nambiar	&	1390	&	Simon Abkarian	&	0.00184	&	Whoopi Goldberg		&	3.057	\\
99	&	Urvashi	&	1389	&	Orson Welles	&	0.00183	&	Dean Stockwell		&	3.057	\\
100	&	Salim Kumar	&	1388	&	Irrfan Khan	&	0.00183	&	Jackie Chan		&	3.058	\\
\hline
\end{longtable}

\section{Centrality Analysis by Decade}
\label{app:CentAnalDecade}

The following table lists the top twenty actors in movies released by decade according to degree centrality, closeness centrality and betweenness centrality. Once again, note the presence of a small number of strange entries in this tables, such as actors with the single names ``Art'', ``Joseph'', ``David'' and ``John''.
\begin{longtable}{|l|ll|ll|ll|} 
\hline\hline
\multicolumn{1}{|}{} &
\multicolumn{2}{c}{Degree Centrality} &
\multicolumn{2}{c}{} &
\multicolumn{2}{c|}{Closeness Centrality} \\
\multicolumn{1}{|}{} &
\multicolumn{2}{c}{(Num. Acting Collaborations)} &
\multicolumn{2}{c}{Betweenness Centrality} &
\multicolumn{2}{c|}{(Mean Path Length)} \\
\hline
\#	& Actor & Score & Actor & Score & Actor & Score \\
\hline
\multicolumn{7}{|l|}{\emph{1910--1920 (single connected network with 9,282 nodes (actors) and 109,181 edges.)}}\\
1	&	Mary Pickford	&	353	&	Anna Jordan	&	0.14071	&	Mary Pickford	&	2.859	\\
2	&	Harold Lloyd	&	327	&	Lottie Pickford	&	0.06066	&	James Neill	&	2.896	\\
3	&	Herbert Standing	&	308	&	Ludwig Trautmann	&	0.04839	&	Tully Marshall	&	2.901	\\
4	&	Harry Carey	&	307	&	Asta Nielsen	&	0.03744	&	Edythe Chapman	&	2.910	\\
5	&	James Neill	&	298	&	Friedrich Kuhne	&	0.03082	&	Herbert Standing	&	2.929	\\
6	&	Spottiswoode Aitken	&	296	&	Charles Villiers	&	0.02735	&	Charles West	&	2.932	\\
7	&	Theodore Roberts	&	293	&	Armand Cortes	&	0.02530	&	Wallace Reid	&	2.933	\\
8	&	Lillian Gish	&	282	&	Arthur Shirley	&	0.02527	&	Owen Moore	&	2.937	\\
9	&	Raymond Hatton	&	280	&	Rita Jolivet	&	0.02235	&	Thomas Meighan	&	2.940	\\
10	&	Lionel Barrymore	&	276	&	George Moss	&	0.02155	&	Jack Pickford	&	2.942	\\
11	&	Lon Chaney	&	275	&	Herbert Standing	&	0.02131	&	Marguerite Clark	&	2.948	\\
12	&	Robert Harron	&	269	&	Winter Hall	&	0.02036	&	Theodore Roberts	&	2.962	\\
13	&	Wallace Reid	&	262	&	Victor Janson	&	0.01816	&	Robert Harron	&	2.963	\\
14	&	Alfred Paget	&	261	&	Gertrude McCoy	&	0.01715	&	Mae Marsh	&	2.965	\\
15	&	Bebe Daniels	&	253	&	Maria Caserini	&	0.01616	&	Raymond Hatton	&	2.980	\\
16	&	Tully Marshall	&	251	&	Lina Cavalieri	&	0.01607	&	Alfred Paget	&	2.982	\\
17	&	Marguerite Clark	&	251	&	Mary Pickford	&	0.01515	&	Irving Cummings	&	2.984	\\
18	&	Walter Long	&	245	&	Marguerite Clark	&	0.01494	&	Donald Crisp	&	2.984	\\
19	&	Bud Jamison	&	245	&	Harry Beaumont	&	0.01406	&	Frank Losee	&	2.990	\\
20	&	William Elmer	&	244	&	Harry Lorraine	&	0.01390	&	Wilfred Lucas	&	2.998	\\
\hline
\multicolumn{7}{|l|}{\emph{1920--1930 (single connected network with 15,519 nodes (actors) and 297,025 edges.)}}\\
1	&	Hermann Picha	&	695	&	Oreste Bilancia	&	0.02072	&	George Fawcett	&	2.628	\\
2	&	George Fawcett	&	618	&	Charles Puffy	&	0.02064	&	Mary Carr	&	2.631	\\
3	&	Karl Platen	&	589	&	Letizia Quaranta	&	0.01821	&	Winter Hall	&	2.644	\\
4	&	Frida Richard	&	587	&	Ivan Koval-Samborsky	&	0.01673	&	Lionel Barrymore	&	2.649	\\
5	&	Wilhelm Diegelmann	&	573	&	Luiza Valle	&	0.01647	&	Charles Puffy	&	2.651	\\
6	&	Louise Fazenda	&	557	&	Igo Sym	&	0.01502	&	Nigel Barrie	&	2.661	\\
7	&	Fritz Kampers	&	547	&	Victor Varconi	&	0.01402	&	Clive Brook	&	2.668	\\
8	&	Tully Marshall	&	527	&	Nato Vachnadze	&	0.01309	&	Edmund Burns	&	2.679	\\
9	&	Lionel Belmore	&	510	&	Augusto Anibal	&	0.01262	&	Gustav von Seyffertitz	&	2.680	\\
10	&	Margarete Kupfer	&	509	&	Henry Victor	&	0.01241	&	Pola Negri	&	2.684	\\
11	&	Lucien Littlefield	&	505	&	Warwick Ward	&	0.01211	&	Anna May Wong	&	2.686	\\
12	&	Maria Forescu	&	499	&	Mary Carr	&	0.01197	&	George Siegmann	&	2.688	\\
13	&	Hans Albers	&	486	&	Sylvia Torf	&	0.01193	&	Evelyn Brent	&	2.689	\\
14	&	Wallace Beery	&	484	&	Nita Naldi	&	0.01152	&	Betty Compson	&	2.689	\\
15	&	Eduard von Winterstein	&	471	&	Lionel Barrymore	&	0.01117	&	Ben Lyon	&	2.692	\\
16	&	Boris Karloff	&	469	&	Vsevolod Pudovkin	&	0.01109	&	Emily Fitzroy	&	2.694	\\
17	&	Lydia Potechina	&	461	&	Nigel Barrie	&	0.01070	&	Warwick Ward	&	2.695	\\
18	&	Julius Falkenstein	&	459	&	Edmund Burns	&	0.01039	&	Norman Kerry	&	2.698	\\
19	&	Josef Swickard	&	455	&	Mary Nolan	&	0.01008	&	Percy Marmont	&	2.700	\\
20	&	Georg John	&	453	&	Pola Negri	&	0.01007	&	Neil Hamilton	&	2.703	\\
\hline
\multicolumn{7}{|l|}{\emph{1930--1940 (single connected network with 22,986 nodes (actors) and 645,435 edges.)}}\\
1	&	Spencer Charters	&	912	&	Charlie	&	0.05117	&	Vladimir Sokoloff	&	2.824	\\
2	&	Ward Bond	&	890	&	Ivan Koval-Samborsky	&	0.04690	&	Peter Lorre	&	2.832	\\
3	&	George Irving	&	888	&	Leonid Kmit	&	0.03409	&	C. Aubrey Smith	&	2.848	\\
4	&	Wade Boteler	&	863	&	Art	&	0.03391	&	Ben Welden	&	2.886	\\
5	&	Irving Bacon	&	850	&	Willy Castello	&	0.02206	&	Herman Bing	&	2.887	\\
6	&	J. Farrell MacDonald	&	824	&	Vladimir Sokoloff	&	0.02132	&	Lilian Harvey	&	2.893	\\
7	&	Paul Hurst	&	811	&	Jan Kiepura	&	0.02053	&	Frank Reicher	&	2.896	\\
8	&	Paul Fix	&	802	&	Bing Crosby	&	0.02012	&	Edward Everett Horton	&	2.898	\\
9	&	Henry Kolker	&	801	&	Heintje Davids	&	0.01787	&	Spencer Charters	&	2.899	\\
10	&	Warren Hymer	&	772	&	Carlos Machado	&	0.01534	&	Reginald Owen	&	2.902	\\
11	&	Paul Horbiger	&	746	&	Peter Lorre	&	0.01414	&	Morgan Wallace	&	2.905	\\
12	&	Willard Robertson	&	738	&	Morgan Wallace	&	0.01350	&	Edmund Lowe	&	2.905	\\
13	&	Oscar Apfel	&	737	&	Sig Arno	&	0.01271	&	Paul Porcasi	&	2.907	\\
14	&	Robert Warwick	&	731	&	Igor Ilyinsky	&	0.01249	&	J. Carrol Naish	&	2.909	\\
15	&	Frank Reicher	&	728	&	Mesquitinha	&	0.01189	&	Henry Kolker	&	2.910	\\
16	&	Berton Churchill	&	725	&	Yakub	&	0.01170	&	Robert Warwick	&	2.912	\\
17	&	Stanley Fields	&	725	&	Papanasam Sivan	&	0.01155	&	Paul Fix	&	2.916	\\
18	&	Joseph Crehan	&	723	&	Conchita Montenegro	&	0.01148	&	Ray Milland	&	2.919	\\
19	&	George Chandler	&	717	&	Tutta Rolf	&	0.01115	&	Victor Varconi	&	2.924	\\
20	&	Karl Platen	&	717	&	Alberto Terrones	&	0.01043	&	Billy Gilbert	&	2.926	\\
\hline
\multicolumn{7}{|l|}{\emph{1940--1950 (single connected network with 23,656 nodes (actors) and 603,812 edges.)}}\\
1	&	Emory Parnell	&	1066	&	Pal	&	0.14073	&	Harry Davenport	&	3.103	\\
2	&	Pierre Watkin	&	971	&	Dev Anand	&	0.09101	&	Alan Napier	&	3.147	\\
3	&	Irving Bacon	&	863	&	Rossano Brazzi	&	0.06714	&	Reginald Owen	&	3.157	\\
4	&	Russell Hicks	&	839	&	Amir Banu	&	0.04972	&	Fortunio Bonanova	&	3.158	\\
5	&	Ray Walker	&	835	&	Arturo de Cordova	&	0.04787	&	Emory Parnell	&	3.159	\\
6	&	Addison Richards	&	753	&	Kamala Kotnis	&	0.04722	&	Frank Puglia	&	3.163	\\
7	&	Joseph Crehan	&	736	&	Hans Klering	&	0.03536	&	Morris Ankrum	&	3.173	\\
8	&	Jerome Cowan	&	734	&	Yelena Tyapkina	&	0.03188	&	Marcel Dalio	&	3.182	\\
9	&	Lloyd Corrigan	&	732	&	Signe Hasso	&	0.02587	&	Gene Lockhart	&	3.183	\\
10	&	John Litel	&	715	&	Marcel Dalio	&	0.02525	&	Akim Tamiroff	&	3.186	\\
11	&	Thurston Hall	&	699	&	Anjali Devi	&	0.02310	&	Peter Lawford	&	3.188	\\
12	&	Chester Clute	&	698	&	Hans Straat	&	0.01624	&	Joseph Cotten	&	3.206	\\
13	&	Byron Foulger	&	688	&	Mai Zetterling	&	0.01623	&	Sig Ruman	&	3.212	\\
14	&	Selmer Jackson	&	687	&	Erich Ponto	&	0.01555	&	Leon Ames	&	3.212	\\
15	&	Richard Lane	&	685	&	Ingrid Bergman	&	0.01444	&	Lloyd Corrigan	&	3.212	\\
16	&	Edward Gargan	&	676	&	Florence Marly	&	0.01441	&	Mikhail Rasumny	&	3.215	\\
17	&	Roy Barcroft	&	658	&	Fortunio Bonanova	&	0.01406	&	Lloyd Nolan	&	3.219	\\
18	&	Douglas Fowley	&	652	&	Pushpavalli	&	0.01351	&	Ray Walker	&	3.219	\\
19	&	Charles Halton	&	649	&	Kishore Sahu	&	0.01330	&	Jerome Cowan	&	3.220	\\
20	&	Harry Davenport	&	642	&	Virgilio Teixeira	&	0.01277	&	Matthew Boulton	&	3.221	\\
\hline
\multicolumn{7}{|l|}{\emph{1950--1960 (single connected network with 36,549 nodes (actors) and 786,464 edges.)}}\\
1	&	Louis de Funes	&	1110	&	George Thorpe	&	0.04861	&	Peter van Eyck	&	2.970	\\
2	&	Sam Kydd	&	773	&	Helen	&	0.03457	&	Yvonne De Carlo	&	3.005	\\
3	&	Sid James	&	755	&	David	&	0.02746	&	Orson Welles	&	3.009	\\
4	&	Richard Wattis	&	685	&	I. S. Johar	&	0.02325	&	Dawn Addams	&	3.009	\\
5	&	Eric Pohlmann	&	650	&	Dora Bryan	&	0.02036	&	Herbert Lom	&	3.014	\\
6	&	Emory Parnell	&	645	&	Carlos Thompson	&	0.01939	&	Sid James	&	3.018	\\
7	&	Morris Ankrum	&	607	&	Padmini	&	0.01665	&	Finlay Currie	&	3.020	\\
8	&	Geoffrey Keen	&	578	&	Nadira	&	0.01647	&	Eric Pohlmann	&	3.026	\\
9	&	Hans Leibelt	&	553	&	Nikolai Kryuchkov	&	0.01635	&	Richard Wattis	&	3.038	\\
10	&	Martin Boddey	&	535	&	Louis de Funes	&	0.01549	&	Anthony Quinn	&	3.039	\\
11	&	Marianne Stone	&	534	&	Sonja Ziemann	&	0.01482	&	Milly Vitale	&	3.040	\\
12	&	Michael Ripper	&	532	&	Nicole Maurey	&	0.01405	&	Patricia Medina	&	3.043	\\
13	&	Ernst Waldow	&	530	&	Raymond Burr	&	0.01391	&	Gina Lollobrigida	&	3.048	\\
14	&	Reinhard Kolldehoff	&	528	&	Marvin Miller	&	0.01387	&	Jack Lambert	&	3.052	\\
15	&	Cyril Chamberlain	&	526	&	William Holden	&	0.01203	&	Raymond Burr	&	3.057	\\
16	&	Myron Healey	&	525	&	Peter van Eyck	&	0.01116	&	Rock Hudson	&	3.058	\\
17	&	Whit Bissell	&	524	&	Steve Cochran	&	0.01085	&	Fernandel	&	3.061	\\
18	&	Nerio Bernardi	&	513	&	Johnson	&	0.01071	&	Mario Siletti	&	3.067	\\
19	&	Lyle Talbot	&	509	&	Willi Narloch	&	0.01069	&	Kirk Douglas	&	3.069	\\
20	&	Laurence Naismith	&	507	&	Ada Vojtsik	&	0.01043	&	Victor McLaglen	&	3.071	\\
\hline
\multicolumn{7}{|l|}{\emph{1960--1970 (single connected network with 39,217 nodes (actors) and 704,331 edges.)}}\\
1	&	Nagesh	&	700	&	William Kerwin	&	0.00343	&	Sylva Koscina	&	2.884	\\
2	&	John Le Mesurier	&	654	&	Lando Buzzanca	&	0.00341	&	Klaus Kinski	&	2.890	\\
3	&	Klaus Kinski	&	639	&	Brian Keith	&	0.00340	&	Senta Berger	&	2.935	\\
4	&	Manorama	&	619	&	Peter Finch	&	0.00338	&	Orson Welles	&	2.937	\\
5	&	Terry-Thomas	&	605	&	Glenn Ford	&	0.00336	&	Terry-Thomas	&	2.940	\\
6	&	David Lodge	&	588	&	Dean Martin	&	0.00332	&	Adolfo Celi	&	2.947	\\
7	&	Warren Mitchell	&	582	&	Juan Carlos Galvan	&	0.00332	&	Robert Morley	&	2.950	\\
8	&	Marianne Stone	&	574	&	Curd Jurgens	&	0.00329	&	Harry Andrews	&	2.956	\\
9	&	John Wayne	&	561	&	Trevor Howard	&	0.00329	&	Wolfgang Preiss	&	2.960	\\
10	&	Narasimharaju	&	561	&	Claudine Auger	&	0.00326	&	Gert Frobe	&	2.968	\\
11	&	Jayanthi	&	554	&	Walter Pidgeon	&	0.00325	&	Mark Damon	&	2.974	\\
12	&	Balakrishna	&	540	&	Philippe Noiret	&	0.00324	&	Curd Jurgens	&	2.980	\\
13	&	Rajkumar	&	539	&	Kirk Douglas	&	0.00323	&	Akim Tamiroff	&	2.985	\\
14	&	Jean-Paul Belmondo	&	534	&	Margaret Lee	&	0.00319	&	Elke Sommer	&	2.987	\\
15	&	Fernando Sancho	&	515	&	Lee Pang-fei	&	0.00319	&	Norman Rossington	&	2.992	\\
16	&	Sylva Koscina	&	508	&	Marne Maitland	&	0.00319	&	Sergio Fantoni	&	2.996	\\
17	&	Robert Morley	&	503	&	Jagdev	&	0.00318	&	David Niven	&	2.997	\\
18	&	Kirk Douglas	&	492	&	Merry Anders	&	0.00315	&	Yul Brynner	&	2.999	\\
19	&	Graham Stark	&	489	&	Chan Wai-yue	&	0.00311	&	Henry Fonda	&	2.999	\\
20	&	Richard Wattis	&	489	&	Tito Garcia	&	0.00308	&	John Wayne	&	3.002	\\
\hline
\multicolumn{7}{|l|}{\emph{1970--1980 (single connected network with 47,321 nodes (actors) and 751,203 edges.)}}\\
1	&	Manorama	&	691	&	John Saxon	&	0.03881	&	Adolfo Celi	&	3.116	\\
2	&	Sankaradi	&	674	&	I. S. Johar	&	0.03065	&	Harry Andrews	&	3.120	\\
3	&	Adoor Bhasi	&	668	&	Kabir Bedi	&	0.02926	&	Michael Caine	&	3.136	\\
4	&	Bahadoor	&	617	&	Richard Attenborough	&	0.02461	&	Donald Pleasence	&	3.140	\\
5	&	Lakshmi	&	556	&	Ivo Garrani	&	0.02299	&	Denholm Elliott	&	3.147	\\
6	&	Helen	&	551	&	Joseph	&	0.01756	&	John Saxon	&	3.160	\\
7	&	Jayabharathi	&	521	&	Dharmendra	&	0.01548	&	George Kennedy	&	3.165	\\
8	&	Prem Nazir	&	514	&	Adolfo Celi	&	0.01423	&	Christopher Lee	&	3.166	\\
9	&	Klaus Kinski	&	508	&	Peter Jones	&	0.01339	&	Martin Balsam	&	3.175	\\
10	&	Balakrishna	&	503	&	Klaus Kinski	&	0.01305	&	Orson Welles	&	3.180	\\
11	&	John Carradine	&	502	&	Geraldine Chaplin	&	0.01221	&	Joseph Cotten	&	3.182	\\
12	&	K. P. Ummer	&	496	&	Donald Pleasence	&	0.01211	&	James Mason	&	3.189	\\
13	&	Ku Feng	&	492	&	Aruna Irani	&	0.01191	&	Elke Sommer	&	3.195	\\
14	&	Sukumari	&	484	&	Harry Andrews	&	0.01126	&	Ray Milland	&	3.197	\\
15	&	Christopher Lee	&	482	&	Oleg Vidov	&	0.01073	&	Donald Sutherland	&	3.203	\\
16	&	Donald Pleasence	&	480	&	Sylvia Miles	&	0.01027	&	Richard Attenborough	&	3.208	\\
17	&	Manjula	&	475	&	Sanjeev Kumar	&	0.01009	&	Fernando Rey	&	3.210	\\
18	&	Nagesh	&	470	&	Christopher Lee	&	0.00975	&	Elliott Gould	&	3.211	\\
19	&	Denholm Elliott	&	466	&	George Kennedy	&	0.00967	&	John Carradine	&	3.213	\\
20	&	Hsu Hsia	&	465	&	William Thomas	&	0.00955	&	Gene Hackman	&	3.218	\\
\hline
\multicolumn{7}{|l|}{\emph{1980--1990 (single connected network with 50,628 nodes (actors) and 776,358 edges.)}}\\
1	&	Sukumari	&	766	&	Amrish Puri	&	0.04610	&	Martin Sheen	&	3.188	\\
2	&	Ambika	&	765	&	Saeed Jaffrey	&	0.02497	&	John Hurt	&	3.217	\\
3	&	Anuradha	&	752	&	Roy Chiao	&	0.01607	&	John Gielgud	&	3.221	\\
4	&	Jayamalini	&	722	&	Martin Sheen	&	0.01461	&	Max von Sydow	&	3.238	\\
5	&	Mithun Chakraborty	&	665	&	Tom Alter	&	0.01454	&	Richard Griffiths	&	3.245	\\
6	&	Seema	&	663	&	Roy Kinnear	&	0.01304	&	Denholm Elliott	&	3.250	\\
7	&	Radha	&	654	&	Janaki	&	0.01228	&	Kenneth McMillan	&	3.262	\\
8	&	Manorama	&	649	&	John Gielgud	&	0.01200	&	Christopher Lee	&	3.269	\\
9	&	Silk Smitha	&	640	&	James Fox	&	0.01159	&	Harvey Keitel	&	3.274	\\
10	&	Vishnuvardhan	&	628	&	Christopher Lee	&	0.01150	&	Michael Hordern	&	3.284	\\
11	&	Geetha	&	623	&	Max von Sydow	&	0.01146	&	M. Emmet Walsh	&	3.289	\\
12	&	Jagathy Sreekumar	&	622	&	Klaus Kinski	&	0.01095	&	Oliver Reed	&	3.289	\\
13	&	Shakti Kapoor	&	620	&	John Hurt	&	0.01095	&	Donald Pleasence	&	3.291	\\
14	&	Srividya	&	602	&	Isabelle Huppert	&	0.01015	&	Robert Loggia	&	3.296	\\
15	&	Mammootty	&	593	&	Jackie Chan	&	0.00995	&	Klaus Kinski	&	3.299	\\
16	&	Nedumudi Venu	&	583	&	Klaus Maria Brandauer	&	0.00967	&	Peter Boyle	&	3.306	\\
17	&	Madhavi	&	578	&	Igor Yasulovich	&	0.00944	&	Harry Dean Stanton	&	3.307	\\
18	&	Amrish Puri	&	570	&	Marcello Mastroianni	&	0.00915	&	Roy Kinnear	&	3.312	\\
19	&	Andy Lau	&	566	&	Everett McGill	&	0.00886	&	Shane Rimmer	&	3.313	\\
20	&	Rekha	&	565	&	Rohini Hattangadi	&	0.00849	&	Edward Fox	&	3.315	\\
\hline
\multicolumn{7}{|l|}{\emph{1990--2000 (single connected network with 56,962 nodes (actors) and 966,791 edges.)}}\\
1	&	Vijayakumar	&	1065	&	Om Puri	&	0.03197	&	Samuel L. Jackson	&	3.067	\\
2	&	Senthil	&	1038	&	Roshan Seth	&	0.03107	&	Frank Welker	&	3.102	\\
3	&	Goundamani	&	865	&	Shabana Azmi	&	0.01726	&	Greta Scacchi	&	3.123	\\
4	&	Srividya	&	857	&	Suman	&	0.01631	&	Whoopi Goldberg	&	3.123	\\
5	&	Brahmanandam	&	815	&	Danny Denzongpa	&	0.01532	&	John Cusack	&	3.124	\\
6	&	Nassar	&	806	&	Greta Scacchi	&	0.01502	&	Julianne Moore	&	3.147	\\
7	&	Murali	&	802	&	Maggie Cheung	&	0.01450	&	Christopher Walken	&	3.150	\\
8	&	Frank Welker	&	796	&	Nagma	&	0.01419	&	Stephen Tobolowsky	&	3.152	\\
9	&	Andy Lau	&	794	&	Shakti Kapoor	&	0.01408	&	Paul Guilfoyle	&	3.158	\\
10	&	Mithun Chakraborty	&	781	&	Captain Raju	&	0.01401	&	James Earl Jones	&	3.160	\\
11	&	Venniradai Moorthy	&	781	&	Gulshan Grover	&	0.01072	&	Mike Starr	&	3.161	\\
12	&	Jagathy Sreekumar	&	772	&	Stellan Skarsgard	&	0.01057	&	Dan Hedaya	&	3.165	\\
13	&	Shakti Kapoor	&	763	&	Valeria Golino	&	0.00985	&	Val Kilmer	&	3.166	\\
14	&	Sukumari	&	755	&	Frank Welker	&	0.00915	&	John Turturro	&	3.170	\\
15	&	Meena	&	751	&	Shashi Kapoor	&	0.00908	&	Denzel Washington	&	3.173	\\
16	&	Manorama	&	732	&	King Kong	&	0.00847	&	Anthony Hopkins	&	3.178	\\
17	&	Thilakan	&	726	&	Rekha	&	0.00813	&	Richard E. Grant	&	3.179	\\
18	&	Delhi Ganesh	&	709	&	Tcheky Karyo	&	0.00787	&	Bruce Willis	&	3.179	\\
19	&	Charle	&	696	&	Michael Ironside	&	0.00736	&	Denis Leary	&	3.180	\\
20	&	Rekha	&	683	&	Pete Postlethwaite	&	0.00720	&	David Thewlis	&	3.181	\\
\hline
\multicolumn{7}{|l|}{\emph{2000--2010 (single connected network with 91,150 nodes (actors) and 1,441,852 edges.)}}\\
1	&	Brahmanandam	&	1307	&	Om Puri	&	0.03160	&	Angelina Jolie	&	3.135	\\
2	&	Nassar	&	1129	&	Jackie Chan	&	0.02348	&	Samuel L. Jackson	&	3.142	\\
3	&	Prakash Raj	&	1108	&	David Carradine	&	0.01230	&	David Carradine	&	3.145	\\
4	&	Ashish Vidyarthi	&	1022	&	Anupam Kher	&	0.01108	&	Keith David	&	3.152	\\
5	&	Tanikella Bharani	&	1008	&	Snoop Dogg	&	0.00958	&	Michael Madsen	&	3.158	\\
6	&	Ali	&	987	&	Ashish Vidyarthi	&	0.00864	&	Jackie Chan	&	3.160	\\
7	&	Jagathy Sreekumar	&	973	&	Juliette Binoche	&	0.00738	&	Harvey Keitel	&	3.161	\\
8	&	Venu Madhav	&	921	&	Monica	&	0.00732	&	Brian Cox	&	3.162	\\
9	&	Sunil	&	910	&	Naseeruddin Shah	&	0.00682	&	Owen Wilson	&	3.171	\\
10	&	M. S. Narayana	&	887	&	Gerard Depardieu	&	0.00641	&	Snoop Dogg	&	3.177	\\
11	&	Kota Srinivasa Rao	&	887	&	Irrfan Khan	&	0.00612	&	Steve Coogan	&	3.177	\\
12	&	Kalabhavan Mani	&	876	&	Stellan Skarsgard	&	0.00580	&	Steve Buscemi	&	3.181	\\
13	&	Vivek	&	858	&	Milind Soman	&	0.00573	&	John Cleese	&	3.185	\\
14	&	Vijayakumar	&	833	&	Nassar	&	0.00558	&	Danny Trejo	&	3.186	\\
15	&	Vadivelu	&	825	&	Sonu Sood	&	0.00548	&	Morgan Freeman	&	3.186	\\
16	&	Dharmavarapu Subramanyam	&	811	&	Eriq Ebouaney	&	0.00545	&	Peter Stormare	&	3.189	\\
17	&	Cochin Haneefa	&	805	&	Anthony Wong	&	0.00536	&	Willem Dafoe	&	3.189	\\
18	&	Devan	&	801	&	Gulshan Grover	&	0.00529	&	Luke Wilson	&	3.190	\\
19	&	Om Puri	&	761	&	Thomas Kretschmann	&	0.00528	&	Woody Harrelson	&	3.190	\\
20	&	Salim Kumar	&	759	&	Udo Kier	&	0.00525	&	Thomas Kretschmann	&	3.191	\\
\hline
\multicolumn{7}{|l|}{\emph{2010--2020 (single connected network with 108,739 nodes (actors) and 1,756,677 edges.)}}\\
1	&	Nassar	&	1689	&	Nassar	&	0.02662	&	Ben Kingsley	&	3.151	\\
2	&	Manobala	&	1314	&	Anupam Kher	&	0.02412	&	Shea Whigham	&	3.170	\\
3	&	Prakash Raj	&	1312	&	Om Puri	&	0.01406	&	Liam Neeson	&	3.180	\\
4	&	Jayaprakash	&	1101	&	Adil Hussain	&	0.01301	&	Naomi Watts	&	3.185	\\
5	&	Brahmanandam	&	1099	&	Isabelle Huppert	&	0.00841	&	Anupam Kher	&	3.192	\\
6	&	Indrans	&	1001	&	Peter Stormare	&	0.00789	&	Charles Dance	&	3.199	\\
7	&	Siddique	&	957	&	Sam Anderson	&	0.00752	&	Fred Tatasciore	&	3.201	\\
8	&	Ashish Vidyarthi	&	948	&	Ben Kingsley	&	0.00726	&	Alfred Molina	&	3.205	\\
9	&	Ajay	&	925	&	Sridhar	&	0.00694	&	Toby Jones	&	3.208	\\
10	&	Avinash	&	914	&	Irrfan Khan	&	0.00688	&	Danny Huston	&	3.211	\\
11	&	Sayaji Shinde	&	901	&	Liam Neeson	&	0.00677	&	Samuel L. Jackson	&	3.213	\\
12	&	Thambi Ramaiah	&	899	&	Charles Dance	&	0.00635	&	Willem Dafoe	&	3.213	\\
13	&	Ali	&	877	&	Vinnie Jones	&	0.00627	&	Antonio Banderas	&	3.213	\\
14	&	Rajendran	&	859	&	Joy Badlani	&	0.00626	&	James Franco	&	3.223	\\
15	&	Nedumudi Venu	&	848	&	Boman Irani	&	0.00609	&	Mark Strong	&	3.227	\\
16	&	Vijayaraghavan	&	845	&	Eric Roberts	&	0.00596	&	Vinnie Jones	&	3.228	\\
17	&	Vennela Kishore	&	844	&	Danny Trejo	&	0.00592	&	Tom Wilkinson	&	3.230	\\
18	&	Lena	&	837	&	Sanjay Mishra	&	0.00585	&	Danny Trejo	&	3.241	\\
19	&	Sunil Sukhada	&	835	&	Willem Dafoe	&	0.00585	&	Stephen Lang	&	3.241	\\
20	&	Tanikella Bharani	&	831	&	John	&	0.00583	&	Christopher Lloyd	&	3.244	\\
\hline
\end{longtable}

%---------------------------------------------------------------------------
\end{document}